\documentclass[onecolumn,10pt]{IEEEtran}
\usepackage[mathscr]{eucal}
\usepackage[cmex10]{amsmath}
\usepackage{epsfig,epsf,psfrag}
\usepackage{amssymb,amsmath,amsthm,amsfonts,latexsym}
\usepackage{amsmath,graphicx,bm,xcolor,url,overpic}
\usepackage{fixltx2e}
\usepackage{array}
\usepackage{verbatim}
\usepackage{bm}
\usepackage{algorithmic}
\usepackage{algorithm}
\usepackage{verbatim}
\usepackage{textcomp}
\usepackage{mathrsfs}
\usepackage{epstopdf}

\newcommand{\openone}{\leavevmode\hbox{\small1\normalsize\kern-.33em1}}

\catcode`~=11 \def\UrlSpecials{\do\~{\kern -.15em\lower .7ex\hbox{~}\kern .04em}} \catcode`~=13 

\allowdisplaybreaks[4]

\newcommand{\nn}{\nonumber}

\newcommand{\calA}{\mathcal{A}}
\newcommand{\calB}{\mathcal{B}}

\newcommand{\calD}{\mathcal{D}}
\newcommand{\calE}{\mathcal{E}}
\newcommand{\calF}{\mathcal{F}}

\newcommand{\calM}{\mathcal{M}}
\newcommand{\calN}{\mathcal{N}}

\newcommand{\calP}{\mathcal{P}}

\newcommand{\calS}{\mathcal{S}}
\newcommand{\calT}{\mathcal{T}}
\newcommand{\calU}{\mathcal{U}}
\newcommand{\calV}{\mathcal{V}}

\newcommand{\calX}{\mathcal{X}}
\newcommand{\calY}{\mathcal{Y}}
\newcommand{\calZ}{\mathcal{Z}}

\newcommand{\ba}{\mathbf{a}}

\newcommand{\bn}{\mathbf{n}}

\newcommand{\bs}{\mathbf{s}}
\newcommand{\bS}{\mathbf{S}}
\newcommand{\bt}{\mathbf{t}}

\newcommand{\bv}{\mathbf{v}}
\newcommand{\bV}{\mathbf{V}}
\newcommand{\bw}{\mathbf{w}}
\newcommand{\bW}{\mathbf{W}}
\newcommand{\bx}{\mathbf{x}}
\newcommand{\bX}{\mathbf{X}}

\newcommand{\rmc}{\mathrm{c}}

\newcommand{\rmd}{\mathrm{d}}

\newcommand{\rme}{\mathrm{e}}

\newcommand{\rmG}{\mathrm{G}}

\newcommand{\rml}{\mathrm{l}}

\newcommand{\rmP}{\mathrm{P}}
\newcommand{\rmq}{\mathrm{q}}

\newcommand{\rms}{\mathrm{s}}

\newcommand{\rmT}{\mathrm{T}}

\newcommand{\rmv}{\mathrm{v}}
\newcommand{\rmV}{\mathrm{V}}

\newcommand{\bbN}{\mathbb{N}}

\newcommand{\bbR}{\mathbb{R}}

\DeclareMathAlphabet{\mathbsf}{OT1}{cmss}{bx}{n}
\DeclareMathAlphabet{\mathssf}{OT1}{cmss}{m}{sl}

\DeclareSymbolFont{bsfletters}{OT1}{cmss}{bx}{n}  
\DeclareSymbolFont{ssfletters}{OT1}{cmss}{m}{n}
\DeclareMathSymbol{\bsfGamma}{0}{bsfletters}{'000}
\DeclareMathSymbol{\ssfGamma}{0}{ssfletters}{'000}
\DeclareMathSymbol{\bsfDelta}{0}{bsfletters}{'001}
\DeclareMathSymbol{\ssfDelta}{0}{ssfletters}{'001}
\DeclareMathSymbol{\bsfTheta}{0}{bsfletters}{'002}
\DeclareMathSymbol{\ssfTheta}{0}{ssfletters}{'002}
\DeclareMathSymbol{\bsfLambda}{0}{bsfletters}{'003}
\DeclareMathSymbol{\ssfLambda}{0}{ssfletters}{'003}
\DeclareMathSymbol{\bsfXi}{0}{bsfletters}{'004}
\DeclareMathSymbol{\ssfXi}{0}{ssfletters}{'004}
\DeclareMathSymbol{\bsfPi}{0}{bsfletters}{'005}
\DeclareMathSymbol{\ssfPi}{0}{ssfletters}{'005}
\DeclareMathSymbol{\bsfSigma}{0}{bsfletters}{'006}
\DeclareMathSymbol{\ssfSigma}{0}{ssfletters}{'006}
\DeclareMathSymbol{\bsfUpsilon}{0}{bsfletters}{'007}
\DeclareMathSymbol{\ssfUpsilon}{0}{ssfletters}{'007}
\DeclareMathSymbol{\bsfPhi}{0}{bsfletters}{'010}
\DeclareMathSymbol{\ssfPhi}{0}{ssfletters}{'010}
\DeclareMathSymbol{\bsfPsi}{0}{bsfletters}{'011}
\DeclareMathSymbol{\ssfPsi}{0}{ssfletters}{'011}
\DeclareMathSymbol{\bsfOmega}{0}{bsfletters}{'012}
\DeclareMathSymbol{\ssfOmega}{0}{ssfletters}{'012}

\newcommand{\hatS}{\hat{S}}

\newcommand{\hatV}{\hat{V}}

\newcommand{\hatW}{\hat{W}}

\newcommand{\bars}{\bar{s}}

\newcommand{\barv}{\bar{v}}

\newcommand{\barx}{\bar{x}}

\newcommand{\barM}{\bar{M}}

\newcommand{\barX}{\bar{X}}

\DeclareMathOperator*{\argmax}{arg\,max}
\DeclareMathOperator*{\argmin}{arg\,min}

\newtheorem{theorem}{Theorem} 
\newtheorem{lemma}{Lemma}

\newtheorem{corollary}{Corollary}

\newtheorem{definition}{Definition}

\usepackage[utf8]{inputenc} 
\usepackage[T1]{fontenc}    
\usepackage{url,bbm}            
\usepackage{booktabs}       
\usepackage{amsfonts,algorithm,algorithmic}       
\usepackage{nicefrac}       
\usepackage{microtype}      

\usepackage{cite}
\usepackage{subfigure,transparent,color,graphicx} 

\allowdisplaybreaks[2]
\usepackage[ colorlinks = true,
             linkcolor = blue,
             urlcolor  = blue,
             citecolor = red,
             anchorcolor = green,]{hyperref}

\newcommand{\bbo}{\mathbbm{1}}

\begin{document}

\title{Resolution Limits of Non-Adaptive 20 Questions Search for a Moving Target}
\author{Lin Zhou and Alfred Hero \\

\thanks{This paper was partially presented at ICASSP 2021~\cite{zhouicassp2020}.}
\thanks{Lin Zhou is with the School of Cyber Science and Technology, Beihang University, Beijing 100191, China (Email: lzhou@buaa.edu.cn). Alfred Hero is with the Department of Electrical Engineering and Computer Science, University of Michigan, Ann Arbor, MI, USA, 48109-2122 (Email: hero@eecs.umich.edu.).}
\thanks{L. Zhou was partially supported by the National Key Research and Development Program of China under Grant 2020YFB1804800, the National Natural Science Foundation of China under Grants 62201022 and U22B2008, the Beijing Municipal Natural Science Foundation under Grant 4232007. A.O. Hero was partially supported by ARO under grant W911NF1910269, and NNSA under grant DE-NA0003921.}
}
\maketitle

\begin{abstract}
Using the 20 questions estimation framework with query-dependent noise, we study non-adaptive search strategies for a moving target over the unit cube with unknown initial location and velocities under a piecewise constant velocity model. In this search problem, there is an oracle who knows the instantaneous location of the target at any time. Our task is to query the oracle as few times as possible to accurately estimate the location of the target at any specified time. We first study the case where the oracle's answer to each query is corrupted by discrete noise and then generalize our results to the case of additive white Gaussian noise. In our formulation, the performance criterion is the resolution, which is defined as the maximal $L_\infty$ distance between the true locations and estimated locations. We characterize the minimal resolution of an optimal non-adaptive query procedure with a finite number of queries by deriving non-asymptotic and asymptotic bounds. Our bounds are tight in the first-order asymptotic sense when the number of queries satisfies a certain condition and our bounds are tight in the stronger second-order asymptotic sense when the target moves with a constant velocity. To prove our results, we relate the current problem to channel coding, borrow ideas from finite blocklength information theory and construct bounds on the number of possible quantized target trajectories.
\end{abstract}


\section{Introduction}

Consider the problem of search for a moving target with unknown initial location and velocities. This problem arises in many areas, including search and rescue (MH370 airplane), security (botnets) and epidemiology (pathogen tracing). The crux is to design efficient strategies to accurately locate the target as soon as possible to minimize cost and potentially save lives. Since the target velocity and location are unknown, such a search problem is challenging and exhaustive search is infeasible for large search regions. To shed light on the design of efficient search strategies and advance the understanding of best theoretically attainable performance, we mathematically model the search process as a noisy 20 questions problem and derive bounds on minimal achievable resolution using finite blocklength information theory~\cite{polyanskiy2010finite,TanBook,ZhouBook}. 

The noisy 20 questions estimation problem is so named in analogy with the old querying-and-answering game called ``20 questions''~\cite{wiki20question} between two players. The game starts with a player, the ``oracle'', who chooses a question and a secret answer. The task of the other player, the ``questioner'', is to figure out the secret using at most 20 questions. As a social game, the 20 questions oracle is assumed to never lie and always provide true answers, resulting in a noiseless 20 questions problem. Motivated by this social game, R\'enyi~\cite{renyi1961problem} proposed a mathematical formulation of the guessing problem by modeling the secret as a random number that takes values in a finite set and assuming that the oracle could provide an untruthful answer to each query with a certain probability. Ulam~\cite{ulam1991adventures} revisited the problem but assumed that the number of untruthful answers to queries is finite.  The mathematical problem of 20 questions estimation is then known as a Ulam-R\'enyi game and the central focus is to design query procedures that exactly guess the secret using as few queries as possible.

The Ulam-R\'enyi game is equivalent to a search problem for a single one-dimensional stationary target with unknown location over the unit interval. Specifically, the secret is the target location, the oracle is mother nature that knows the secret, the questioner is the designer of a search strategy, who poses a sequence queries to the oracle and obtains noisy answers from the oracle about the presence of the target. There are two types of search strategies: non-adaptive search and adaptive search. In adaptive search the design of the current query depends on all the previous queries and the answers to these queries. In non-adaptive search the sequence of queries is determined before the start of the game. Readers can refer to \cite[Section I]{zhou2019twentyq} for  more details. We focus on the non-adaptive search strategy since it can be applied in an time-efficient parallel manner without the need for a feedback channel.

The Ulam-R\'enyi game has motivated diverse applications including fault-tolerant communications~\cite{renyi1961problem}, human-in-the-loop decision-making~\cite{tsiligkaridis2014collaborative} and object localization in an image~\cite{jedynak2012twenty,rajan2015bayesian}. For the Ulam game, optimal search strategies and corresponding sample complexity were summarized by Pelc in \cite{pelc2002searching}. In contrast, the R\'enyi game is less well understood and has attracted much recent attention~\cite{jedynak2012twenty,chiu2016sequential,tsiligkaridis2014collaborative,tsiligkaridis2015decentralized,kaspi2018searching,chung2018unequal,variani2015non,rajan2015bayesian}.  In these works, the secret location of the target is assumed to be a continuous random variable that takes values in the unit interval and the behavior of the oracle (such as lying or refusing to answer a query) is modeled by a noisy channel. The channel is a probability transition matrix from a binary value (``yes'' or ``no'') to an output alphabet (either continuous or discrete). Among all these works, the setting in \cite{kaspi2018searching} is of particular interest. Firstly, the authors of \cite{kaspi2018searching} proposed a query-dependent noise model where the probability that the oracle lies depends on the size of the query set. Such a query-dependent noise model is relevant to applications such as biometric sensing, high throughput biological tests or search with sensor networks. Secondly, the authors of \cite{kaspi2018searching} considered the absolute estimation error as the performance criterion, in contrast to the differential entropy criterion adopted in earlier works~\cite{jedynak2012twenty}. 

Most studies on 20 questions estimation focus on stationary targets. The exception includes \cite[Theorem 3]{kaspi2018searching} where the authors assumed a one-dimensional moving target over the unit circle. Therein, the authors derived asymptotic bounds on the performance of an optimal non-adaptive search strategy with infinite number of queries assuming query-dependent Bernoulli noise. In this paper, we generalize \cite[Theorem 3]{kaspi2018searching} to the case of a moving target whose position lies in a multidimensional unit cube. Specifically, at time $t=0$, which is the starting time point that a search procedure begins, the initial location of the target is unknown. Subsequently, the target moves with unknown velocities in $B$ time slots. At each time slot $j\in\{1,\ldots,B\}$, the target moves with a constant velocity and changes velocity at the ending time point $n_j$. The ending time points $(n_1,\ldots,n_B)$ of these $B$ time slots are assumed fixed and known. To search for the moving target, in each time slot, we make a fixed number of queries to an oracle that knows real time location of the target and obtain noisy responses. With the goal to minimize the estimation error of the trajectory of the moving target, we study the minimal achievable resolution defined as the $L_{\infty}$ norm of the trajectory estimation error of any non-adaptive query procedure. We consider the query-dependent noise model where the noise is either discrete noise or additive white Gaussian noise (AWGN).

\subsection{Main Contributions}

Our main contributions are the following. We establish non-asymptotic achievability and converse bounds on the performance of optimal non-adaptive moving target search. We introduce a multi-stage search algorithm that uses random coding to generate queries and uses maximal mutual information density decoding to estimate the trajectory. We obtain first-order and second-order asymptotically tight bounds under mild conditions. We demonstrate that our proposed algorithm attains the non-asymptotic achievability and second-order asymptotic bounds.

To derive the non-asymptotic bounds, we borrow ideas from finite blocklength coding, in particular the random coding union bound~\cite{polyanskiy2010finite} and the change-of-measure technique~\cite{csiszar2011information}, and we generalize previous bounds on the number of trajectories of a moving target~\cite[Theorem 3]{kaspi2018searching}. In our achievability proof, we introduce a non-adaptive query procedure (cf. Algorithm \ref{procedure:nonadapt}) that uses the maximal mutual information density decoder. When the channel is a query-dependent BSC or AWGN channel, the decoder in Algorithm \ref{procedure:nonadapt} is equivalent to a nearest neighbor decoder, which is independent of the channel parameters. Furthermore, we derive a non-asymptotic converse result, which bounds the best possible performance of any non-adaptive query procedure. Our non-asymptotic converse bound holds for any noisy channel and is derived using the relationship between the current problem and the data transmission (channel coding) problem~\cite{cover2012elements}.

Using the non-asymptotic bounds, under the assumption of bounded maximal speed of the target, we derive asymptotic bounds on the achievable resolution of optimal non-adaptive query procedures. Our results are first-order asymptotically tight when the number of queries for different time slots satisfy a mild condition. We show that a cold restart search, where one searches for the target at each time slot ignoring estimation results from previous time slots, is strictly suboptimal even asymptotically. Furthermore, our results are second-order asymptotically tight when the target moves with constant velocity, corresponding to $B=1$. For this case, we establish a phase transition phenomenon: the excess-resolution probability of optimal non-adaptive search exhibits a sharp transition as a function of the resolution decay rate, i.e., the rate of  exponential decay of the resolution with respect to the number of queries. Specifically, if one wishes to achieve a resolution decay rate higher than a critical value, then with probability one, no non-adaptive query procedure can achieve this goal. On the other hand, if one wishes to achieve a resolution decay rate lower than the critical value, then with probability one, one can always achieve the goal with an optimal non-adaptive query procedure, such as Algorithm \ref{procedure:nonadapt}. Note that this sharp transition is analogous to the strong converse theorem for channel coding~\cite{wolfobook}, which asserts a sharp transition of the asymptotic error probability between zero and one as a function of the coding rate. Specifically, as the blocklength increases to infinity, the error probability tends to one if the coding rate is above the capacity while the error probability vanishes if the coding rate is below the capacity. 

\subsection{Comparison with Most Related Works}

The work most closely related to this paper is \cite[Theorem 3]{kaspi2018searching}, \cite{lalitha2018improved} and our previous publications \cite{zhou2019twentyq,zhou2021resolution}. In \cite[Theorem 3]{kaspi2018searching}, the authors initiated the study of search for a moving target with unknown initial location and velocity over the unit circle with query-dependent Bernoulli noise. Under the assumption that the target moves less than half of the circumference of the circle per unit query time, the authors derived bounds on the asymptotic decay rate of the resolution, defined as the  maximal absolute difference between estimated and true values of the location or the velocity. The bounds in \cite[Theorem 3]{kaspi2018searching} are tight when the maximal speed tends to zero, thus establishing the asymptotically optimal resolution decay rate as the number of queries tends to \emph{infinity}. We generalize the setting of \cite[Theorem 3]{kaspi2018searching} by considering a multidimensional target, a piecewise constant velocity model and a more general noise model. Furthermore, we derive i) non-asymptotic achievability and converse bounds and ii) second-order asymptotics that provide approximations to the performance of an optimal non-adaptive query procedure with finitely many queries. Note that non-asymptotic and second-order asymptotic bounds are important since in practical search problems one usually tracks a target for a limited finite amount of time. 

The authors of \cite{lalitha2018improved} generalized the setting of \cite[Theorem 1]{kaspi2018searching} to the case of search for a stationary one-dimensional target with query-dependent AWGN and derived asymptotic resolution decay rates for non-adaptive and adaptive query procedures. Note that in \cite{lalitha2018improved}, the authors focused on adaptive queries and only presented a weak converse result (based on Fano's inequality) for non-adaptive query procedures. In contrast, we study the search for a moving multidimensional target and derive both non-asymptotic and second-order asymptotic results for the same AWGN setting as \cite{lalitha2018improved}. Our study for multidimensional moving targets is a significant generalization of the non-adaptive search for a stationary one-dimensional target in \cite{lalitha2018improved}.

Finally, we discuss the differences between the current paper and our previous publications~\cite{zhou2019twentyq,zhou2021resolution} on non-adaptive search. Our first work in \cite{zhou2019twentyq} studies the search problem of a single stationary target using finite blocklength information theory, demonstrates the performance of optimal non-adaptive query and establishes the fact that search over each dimension of the target is strictly suboptimal from a second-order asymptotic perspective. Our second work~\cite{zhou2021resolution} obtains a multiple target extension of \cite{zhou2019twentyq} using different query design and different proof techniques. The current paper generalizes \cite{zhou2019twentyq} in another direction, addressing the problem of a moving target with piecewise constant velocity. Our analysis uses similar proof techniques as used in \cite{zhou2019twentyq}. However, the case of a moving target makes the analysis more complicated since the location of the target varies over time, introducing combinatorial complexity due to the need to consider the set of all possible quantized trajectories. Furthermore, we consider an AWGN channel, which violates the continuity assumption for a query-dependent channel in~\cite{zhou2019twentyq}. Finally, we bring new intuition beyond~\cite{zhou2019twentyq}: for the piecewise constant velocity model, asymptotically it is strictly suboptimal to do a cold restart search in each time slot.

\subsection{Organization of the Paper}
In Section \ref{sec:pf}, we set up the notation, formulate the problem of search for a moving target and define the fundamental limit of interest. Subsequently, in Section \ref{sec:results}, we present our main results concerning non-asymptotic and second-order asymptotic characterization of the fundamental limit and introduce an optimal search algorithm that achieves the fundamental limit. The proofs of our results are presented in Section \ref{sec:proofs}. Finally, we conclude our paper and discuss future research directions in Section \ref{sec:conclusion}. Most of the proofs are deferred to appendices.

\section{Problem Formulation}
\label{sec:pf}
\subsection*{Notation}
Random variables and their realizations are denoted by upper case variables (e.g.,  $X$) and lower case variables (e.g.,  $x$), respectively. All sets are denoted in calligraphic font (e.g.,  $\mathcal{X}$). Let $X^{n_1}:=(X_1,\ldots,X_n)$ be a random vector of length $n$. We use $\Phi^{-1}(\cdot)$ to denote the inverse of the cumulative distribution function (cdf) of the standard Gaussian density. We use $\bbR$, $\bbR_+$ and $\bbN$ to denote the sets of real numbers, positive real numbers and integers respectively. Given any two integers $(m,n)\in\bbN^2$, we use $[m:n]$ to denote the set of integers $\{m,m+1,\ldots,n\}$ and use $[m]$ to denote $[1:m]$. Given any $(m,n)\in\bbN^2$, for any $m$ by $n$ matrix $\ba=\{a_{i,j}\}_{i\in[m],j\in[n]}$, the infinity norm is defined as $\|\ba\|_{\infty}:=\max_{i\in[m],j\in[n]}|a_{i,j}|$. Given any vector $x^{n_1}$, we use $\|x^{n_1}\|^2$ to denote the $L_2$ norm $\sum_{i\in[n]}x_i^2$. The set of all probability distributions on a finite set $\calX$ is denoted as $\calP(\calX)$ and the set of all conditional probability distributions from $\calX$ to $\calY$ is denoted as $\calP(\calY|\calX)$. Furthermore, we use $\calF(\calS)$ to denote the set of all probability density functions on a set $\calS$. All logarithms are base $e$ unless otherwise noted. Finally, we use $\bbo(\calA)$ to denote the indicator function of an event $\calA$.

\subsection{System Model: 20 Questions Estimation For a Moving Target}
\label{sec:piecewise}
We state the model for the problem of search for a moving target with unknown location and velocities over a unit cube of finite dimension. Fix any finite integer $d\in\bbN$. The initial location of the target is modeled by a $d$-dimensional vector of continuous random variables $\bS=(S_1,\ldots,S_d)$, which take values in $[0,1]^d$. For tractability, we consider a piecewise constant velocity model: the target moves with constant velocity during a time slot and changes its velocity across time slots. Specifically, let $B\in\bbN$ be the total number of time slots and let $(n_1,\ldots,n_B)\in\bbN^B$ be ending time points of slots. For any time point $t\in[n_B]$, the target moves with the velocity $\bV_j=(V_{j,1},\ldots,V_{j,d})\in[-v_+,v_+]^d$ if $t\in[n_{j-1},n_j]$, where we define $n_0:=0$ and $v_+$ is the maximal moving speed per dimension. The collection of velocities at all time slots is denoted by a $B$ by $d$ matrix $\bV^B$, i.e., $\bV^B=(\bV_1,\ldots,\bV_B)\in\calV^{B\times d}$, where for each $j\in[B]$, the $j$-th line corresponds to the $d$-dimensional velocity vector $\bV_j$. The joint distribution $f_{\bS\bV^B}$ of the location vector $\bS$ and the velocity matrix $\bV^B$ is assumed arbitrary and \emph{unknown}.

\begin{figure}[tb]
\centering
\includegraphics[width=.45\columnwidth]{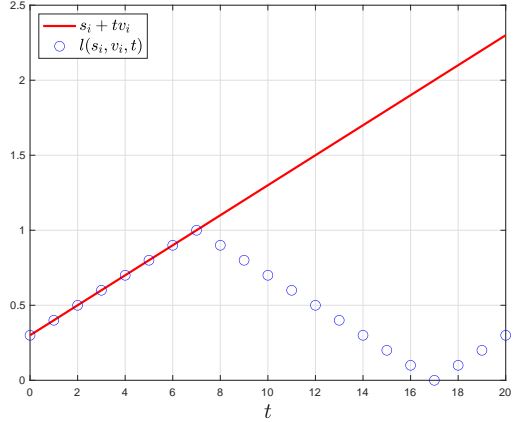}
\caption{Illustration of the location of the target at the $i$-th dimension when $s_i=0.3$ and $v_i=0.1$.}
\label{unitcube}
\end{figure}

Since we consider the unit cube as the feasible region, the location of the target requires clarification. To do so, we first describe the case of the constant velocity model when $B=1$ and then generalize to a piecewise constant velocity model with any finite integer $B$. Specifically, given any $\bs=(s_1,\ldots,s_d)\in[0,1]^d$ and any $\bv=(v_1,\ldots,v_d)\in\calV^d$, if a target has initial location $\bs$ at time $0$ and moves with velocity $\bv$, at any time $t\in\bbN$, for each $i\in[d]$, the $i$-th coordinate $\rml(s_i,v_i,t)$ of the location of the target satisfies
\begin{align}
\rml(s_i,v_i,t)
&:=\left\{
\begin{array}{ll}
s_i+tv_i &\mathrm{if~}\mathrm{mod}(s_i+tv_i,2)\leq 1,\\
2-(s_i+tv_i)&\mathrm{otherwise},
\end{array}
\right.
\label{location}
\end{align}
where $\mathrm{mod}(a,b)$ refers to the modulo operator and outputs the remainder of $a$ dividing $b$. Note that in Eq. \eqref{location}, we take the modulo operator to account for the fact that the target is constrained to move within the unit cube (see Fig. \ref{unitcube} for an illustration at dimension $i$). For the special case of $d=1$, such a setting resembles ~\cite[Theorem 3]{kaspi2018searching} where the target is constrained to move over the unit circle.

Subsequently, the location of the target is generalized to a piecewise constant velocity model. Suppose the initial location of the target is $\bs\in[0,1]^d$ and the velocity of time slot $j\in[B]$ is $\bv_j=(v_{j,1},\ldots,v_{j,d})\in\calV^d$. In the first time slot, at any time $t\in[1,n_1]$, for each $i\in[d]$, the $i$-th coordinate of the location of the target is $\rml(s_i,v_{1,i},t)$ and the collection of target location at all dimensions is denoted as $\rml(\bs,\bv_1,t)$, i.e., $\rml(\bs,\bv_1,t)=(\rml(s_1,v_{1,1},t),\ldots,\rml(s_d,v_{1,d},t))$. For each $j\in[2:B]$, in the $j$-th time slot, at any time $t\in[n_{j-1}+1,n_j]$, for each $i\in[d]$, the $i$-th coordinate of the location of the target  $\rml(s_i,v_i^j,t)$ satisfies $\rml(s_i,v_i^j,t)=\rml(\rml(s_i,v_{j-1,i},n_{j-1}),v_{j,i},t-n_{j-1})$, where $v_i^j=(v_{1,i},\ldots,v_{j,i})$ is the collection of the velocity at dimension $i$ for all time slots until the $j$-th one and $\rml(s_i,v_{j-1,i},n_{j-1})$ is the location of the target at the $i$-th dimension at the end of the previous time slot. The collection of the location of the target in all dimensions for any $t\in[n_{j-1}+1,n_j]$ is similarly denoted as $\rml(\bs,\bv^j,t)=(\rml(s_1,v_1^j,t),\ldots,\rml(s_d,v_d^j,t))$, where $\bv^j=(\bv_1,\ldots,\bv_j)$ is a $j$ by $d$ matrix that collects velocities of the target at all dimensions in the first $j$ time slots.

We remark that our piecewise constant velocity model generalizes the constant velocity setting in~\cite[Section V]{kaspi2018searching} and takes a further step towards practical moving target search. However, our velocity model is still imperfect due to the strong assumptions of knowing the velocity switching times and of an abruptly changing velocity across time slots. The former assumption enables our results to serve as benchmarks for the case of unknown velocity changing times. The latter assumption can be relaxed by controlling the difference of velocities between time slots, e.g.,~assuming that $\max_{j\in[2:B]}\max_{i\in[d]}|V_{j,i}-V_{j-1,i}|\leq\rho v_+$ for some $\rho\in(0,1)$, where $\rho$ controls the acceleration across time slots in each dimension. Our analyses can be generalized to account for this acceleration-constrained case. It will be worthwhile to generalize our velocity model and make it more practical towards real moving target search scenarios, e.g., by considering the velocity as a smooth function of time with bounded derivatives or tolerating random noise and perturbation in the trajectory from piecewise constant velocity models.

We adopt the standard noisy 20 questions estimation framework. There is an oracle who knows the instantaneous locations of the target at any time. Our goal is to pose a sequence of queries $(\calA_1,\ldots,\calA_{n_B})\in([0,1]^d)^{n_B}$ to the oracle and use the oracle's noisy response to accurately estimate the trajectory of the target at any time $t\in[n_B]$. Specifically, at each time point $t\in[n_B]$, the oracle is queried for whether or not the target lies in the region $\calA_t\subseteq[0,1]^d$. The correct answer is $X_t=\sum_{j\in[B]}\bbo\big(t\in[n_{j-1}+1,n_j]\mathrm{~and~}\rml(\bs,\bv_j,t)\in\calA_t\big)$, which is corrupted by a query-dependent noise, modeling the effect of transmitting $X_t$ through a noisy channel. This yields a noisy response $Y_t$ that takes value in an alphabet $\calY$. Given the noisy responses to all $n_B$ queries, we use a decoder $g:\calY^{n_B}\to([0,1]^d)^{n_B}$ to estimate the trajectory of the target.

\subsection{The Query-Dependent Channel}
\label{sec:mdc}
We recall the definition of a query-dependent channel in~\cite{zhou2021resolution}, originally treated in~\cite{kaspi2018searching,chiu2016sequential} and also known as a channel with state~\cite[Chapter 7]{el2011network}. Given any $n\in\bbN$ and any sequence of queries $\calA^{n_1}\subseteq([0,1]^d)^{n_1}$, the channel from the oracle to the player is  memoryless and its transition probabilities are functions of the queries. Specifically, for any $(x^{n_1},y^{n_1})\in\{0,1\}^{n_1}\times\calY^{n_1}$,
\begin{align}
P_{Y^{n_1}|X^{n_1}}^{\calA^{n_1}}(y^{n_1}|x^{n_1})
&=\prod_{t\in[n_1]}P_{Y|X}^{\calA_t}(y_t|x_t),
\end{align}
where $P_{Y|X}^{\calA_t}$ denotes the transition probability of the channel which depends on the $t$-th query $\calA_t$. Given any query $\calA\subseteq[0,1]^d$, let $|\calA|=\int_{\bt\in\calA}\rmd \bt$ be the Lebesgue measure
of $\calA$. The query-dependent channel $P_{Y|X}^{\calA}$ depends on the query $\calA$ only through its size $|\calA|$. Specifically, $P_{Y|X}^{\calA}$ is equivalent to a channel with state $P_{Y|X}^q$, where the state $q$ satisfies $q=f(|\calA|)$ where $f:[0,1]\to\bbR_+$ is a bounded Lipschitz continuous function with parameter $K$, i.e., $|f(q_1)-f(q_2)|\leq K|q_1-q_2|$ and $\max_{q\in[0,1]}f(q)<\infty$. When the distribution of the noise is discrete, the following continuity assumption will be needed. For any $q\in[0,1]$, let $\xi\in(0,\min(q,1-q))$ and assume that there exists a positive constant $c(q)$ depending on $q$ only such that
\begin{align}
\max\left\{\left\|\left\{\log\frac{P_{Y|Z}^q(y,z)}{P_{Y|Z}^{{q+\xi}}(y,z)}\right\}_{(y,z)\in\calY\times\calZ}\right\|_{\infty},\left\|\left\{\log\frac{P_{Y|Z}^q(y,z)}{P_{Y|Z}^{{q-\xi}}(y,z)}\right\}_{(y,z)\in\calY\times\calZ}\right\|_{\infty}\right\}\leq c(q)\xi\label{assump:continuouschannel}.
\end{align}

Two types of query-dependent channels are used in this paper. The first type of the query-dependent channel is discrete with Bernoulli noise.
\begin{definition}
\label{def:mdBSC}
Given any $\calA\subseteq[0,1]$, a channel $P_{Y|X}^{\calA}$ is said to be a query-dependent Binary Symmetric Channel (BSC) with parameter $\zeta\in(0,1]$ if $\calX=\calY=\{0,1\}$ and 
\begin{align}
P_{Y|X}^{\calA}(y|x)=(\zeta f(|\calA|))^{\bbo(y\neq x)}(1-\zeta f(|\calA|))^{\bbo(y=x)},~\forall~(x,y)\in\{-1,1\}^2.
\end{align}
\end{definition}
This channel was introduced in \cite{zhou2021resolution} and generalizes the query-dependent Bernoulli noise model in \cite{kaspi2018searching}, where $\zeta=1$ and $f(|\calA|)=|\calA|$. Note that the output of a query-dependent BSC with parameter $\zeta$ is the input flipped with probability $\zeta f(|\calA|)$. One can verify that the query-dependent BSC satisfies \eqref{assump:continuouschannel}.

The second type of the query-dependent channel is continuous with Gaussian noise. 
\begin{definition}
\label{def:mdAWGN}
Given any $\calA\subseteq[0,1]$, a channel $P_{Y|X}^{\calA}$ is said to be a query-dependent AWGN channel with parameter $\sigma\in\bbR_+$ if $\calX=\{0,1\}$, $\calY=\bbR$ and 
\begin{align}
P_{Y|X}^{\calA}(y|x)=\frac{1}{\sqrt{2\pi(f(|\calA|)\sigma)^2}}\exp\left(-\frac{(y-x)^2}{2(f(|\calA|)\sigma)^2}\right),~(x,y)\in\{0,1\}\times\bbR_+.
\end{align}
\end{definition}
This definition appeared in \cite{lalitha2018improved}. Note that the query-dependent AWGN channel violates the constraint in \eqref{assump:continuouschannel} since $(y-x)^2$ can be unbounded. However, the constraint in Eq. \eqref{assump:continuouschannel} can be made to hold with high probability by restricting $(x,y)$ to satisfy $(y-x)^2<L$, where $L\in\bbR_+$ is arbitrarily large but finite.

These BSC and AWGN channels will be considered to illustrate our results.

\subsection{Definition of the Fundamental Limit}

Recall the piecewise constant velocity model in Section \ref{sec:piecewise} with $B$ different time slots with boundary time points $\bn:=(n_1,\ldots,n_B)$. A non-adaptive 20 questions query procedure is defined as follows.
\begin{definition}
\label{def:procedure:moving}
Given any $\delta\in\bbR_+$ and $\varepsilon\in[0,1]$, an $(\bn,d,\delta,\varepsilon)$-non-adaptive query procedure for moving target search over $[0,1]^d$ consists of 
\begin{itemize}
\item $n_B$ queries $\calA^{n_B}=(\calA_1.\ldots,\calA_{n_B})$, where at each time point $t\in[n_B]$, the oracle is queried whether the target is contained in a Lebesgue measurable set $\calA_t\subseteq[0,1]^d$,
\item an estimator $g:\calY^{n_B}\to[0,1]^d\times\calV^{B\times d}$,
\end{itemize}
such that the excess-resolution probability for the target trajectory satisfies
\begin{align}
\nn&\rmP_\rme(\bn,d,\delta)\\*
&:=\sup_{f_{\bS\bV^B}}\Pr\left\{\max_{t\in[0,n_1]}\|l(\hat{\bS},\hat{\bV}_1,t)-l(\bS,\bV_1,t)\|_{\infty}>\delta\mathrm{~or~}
\max_{j\in[2:B]}\max_{t\in[n_{j-1}+1,n_j]}\|l(\hat{\bS},\hat{\bV}^j,t)-l(\bS,\bV^j,t)\|_{\infty}>\delta\right\}\leq \varepsilon\label{evaluate},
\end{align}
where $\hat{\bS}$ and $\hat{\bV}^B=(\hat{\bV}_1,\ldots,\hat{\bV}_B)$ are the estimated initial location and velocities of the target, respectively.
\end{definition}

Given any $(\bn,d,\delta,\varepsilon)$ query procedure, with probability of at least $1-\varepsilon$, the trajectory of the target can be estimated with resolution $\delta$ at each dimension. When specialized to the case of $d=1$, our formulation in \eqref{evaluate} imposes a slightly stronger requirement than \cite[Theorem 3]{kaspi2018searching}, where the estimates $\hatS$ and $\hatV$ are constrained so that the absolute error between the estimated and true locations of the target at time $t\in\bbN$ is upper bounded by $t\delta$ instead of by $\delta$ in Def. \eqref{def:procedure:moving}. When the maximal speed $v_+=0$, our setting reduces to the search for a stationary target studied in \cite{zhou2019twentyq}.

We remark that constraining the resolution of the estimating the target trajectory is equivalent to constraining the resolution of estimating the initial location and moving velocities $(\bS,\bV)$. Suppose that $\hat{\bS}=(\hatS_1,\ldots,\hatS_d)$ is the estimate of the location vector $\bS=(S_1,\ldots,S_d)$ and $\hat{\bV}^B=(\hat{\bV}_1,\ldots,\hat{\bV}_B)$ is the estimate of the $B$ by $d$ dimensional velocity matrix $\bV=(\bV_1,\ldots,\bV_B)$. For each $j\in[B]$, let $N_j:=n_j-n_{j-1}$. Fix arbitrary $\delta'\in\bbR_+$. It is easy to verify that for each $i\in[d]$, if $|\hatS_i-S_i|<\delta'$ and $\max_{j\in[B]}|\hatV_{j,i}-V_{j,i}|<\frac{\delta'}{N_j}$, no excess-distortion event occurs with respect to the resolution level $\delta=(B+1)\delta'$. On the other hand, if $|\hatS_i-S_i|>\delta$ or $\max_{j\in[B]}|\hatV_{j,i}-V_{j,i}|>\frac{2\delta}{N_j}$, an excess-distortion event with respect to the resolution $\delta$ definitely occurs. Thus, accurately estimating the initial location $\bS$ and moving velocities $\bV^B$ of a target is equivalent to accurately estimating the trajectory $l(\bS,\bV^j,t)$ of the target for each $j\in[B]$ and $t\in[n_{j-1}+1,n_j]$. The above equivalence results are used in the proofs of non-asymptotic achievability and converse results in Theorems \ref{ach:fbl} to \ref{fbl:converse}.

The fundamental limit of interest that characterizes the performance of optimal non-adaptive query procedures is the minimal achievable resolution. Specifically, given any $\bn=(n_1,\ldots,n_B)$, finite dimension $d\in\bbN$ and
tolerable excess-resolution probability $\varepsilon$, the minimal achievable resolution is defined as
\begin{align}
\delta^*(\bn,d,\varepsilon)
&:=\inf\{\delta\in\bbR_+:\exists~\mathrm{an~}(\bn,d,\delta,\varepsilon)\mathrm{-non}\mathrm{-adaptive}\mathrm{~query}\mathrm{~procedure}\}.
\end{align}
We provide non-asymptotic and second-order asymptotic bounds on $\delta^*(\bn,d,\varepsilon)$ for both query-dependent discrete memoryless channels satisfying \eqref{assump:continuouschannel} and for query-dependent AWGN channels.

\section{Main Results}
\label{sec:results}

\subsection{Preliminaries}
\label{sec:preliminaries}
We present necessary preliminary definitions. Given any $(n,M)\in\bbN^2$, define the following quantization function
\begin{align}
\rmq(s,n)
&:=\lceil snM\rceil,~s\in[0,1]\label{quan:move}.
\end{align}
The function $\rmq(s,n)$ will be used to quantize the location of the target so that non-adaptive grid search can  be applied. Recall the definition of the trajectory function $\rml(\cdot)$ around \eqref{location}, the fact that $\bn=(n_1,\ldots,n_B)$ denotes the ending time points of the piecewise velocity model in Section \ref{sec:piecewise} with $B$ time slots and the definition that $n_0=0$. Furthermore, let $N_1=n_1$ and for each $j\in[2:B]$, let $N_j:=n_j-n_{j-1}$. Suppose that a target has location vector $\bs=(s_1,\ldots,s_d)\in[0,1]^d$ at time point zero. Subsequently, for each $j\in[B]$, in the $j$-th time slot $[n_{j-1}+1,n_j]$, the target moves with velocity $\bv_j=(v_{j,1},\ldots,v_{j,d})\in\calV^d$. Then, for each $i\in[d]$ and $j\in[B]$, the quantized location of the target at time $t\in[n_{j-1}+1,n_j]$ along the $i$-th dimension satisfies
\begin{align}
w(s_i,v_i^j,t):=\rmq(l(s_i,v_i^j,t),N_j)\label{quantizei},
\end{align}
and the quantized location of the target satisfies $w(\bs,\bv^j,t)=(w(s_1,v_1^j,t),\ldots,w(s_d,v_d^j,t))$, which is a $d$-dimensional vector.

Note that the trajectory of the target in the first and other time slots are different. This is because in the first time slot, the trajectory of the target is affected by both the initial location $\bs$ and the velocity $\bv_1$ while in the $j$-th time slot for any $j\in[2:B]$, with fixed initial location and velocities of previous time slots, only the velocity $\bv_j$ affects the trajectory. Thus, we define the sets of trajectories of the moving target in different time slots separately. For the first time slot with $n_1$ time points, the set of all possible quantized trajectories is
\begin{align}
\calB_{n_1,M}&:=\big\{\bw^{n_1}\in[n_1M]^{n_1\times d}:\bw^{n_1}=\bw(\bs,\bv_1,[n_1])\mathrm{~for~some~}(\bs,\bv_1)\in[0,1]^d\times\calV^d\big\}
\label{def:calB1},
\end{align}
where $\bw(\bs,\bv_1,[n_1]):=(w(\bs,\bv_1,1),\ldots,w(\bs,\bv_1,n_1))\in[n_1M]^{n_1\times d}$ is the trajectory with initial location $\bs$ and velocity $\bv_1$, i.e., the collection of quantized locations at discrete time points $[n_1]=\{1,\ldots,n_1\}$, of the target. The size of $\calB_{n_1,M}$ is upper bounded by
\begin{align}
|\calB_{n_1,M}|\leq \big(2(n_1v_++3)n_1^4M^2\big)^d\label{upptrajectories}.
\end{align}

The upper bound in \eqref{upptrajectories} is obtained similarly to \cite[Lemma 2]{kaspi2018searching} by considering all possible quantized trajectories at each dimension and taking a upper bound by multiplying the number of trajectories at all dimensions. Specifically, for each $i\in[d]$, let 
\begin{align}
\calB_{n_1,M}^i:=\{w^{n_1}\in[n_M]^{n_1}:~w^{n_1}=(\rml(s_i,v_i,1),\ldots,\rml(s_i,v_i,n_1))~\mathrm{for~some~}(s_i,v_i)\in[0,1]\times\calV\},
\end{align}
and thus $|\calB_{n_1,M}|=\prod_{i\in[d]}|\calB_{n_1,M}^i|$. To upper bound each term $|\calB_{n_1,M}^i|$, similarly to \cite[Lemma 2]{kaspi2018searching}, we need to consider the sets of trajectories $\calB_{n_1,M}^i(t_1,t_2)$ and $\calB_{n_1,M}^{i,\rmc}(t_1,t_2)$ with $(t_1,t_2)\in[n_1M]^2$. The set $\calB_{n_1,M}^i(t_1,t_2)$ collects trajectories whose initial location is in the $t_1$-th interval and whose final location is in the $t_2$-th interval while the set $\calB_{n_1,M}^{i,\rmc}(t_1,t_2)$ is a subset of $\calB_{n_1,M}^i(t_1,t_2)$ with the additional requirement that the target starts and ends at the centers of the $t_1$-th and $t_2$-th intervals, respectively. It follows from \cite[Lemma 2]{kaspi2018searching} that
\begin{align}
|\calB_{n_1,M}^i|
&=\sum_{(t_1,t_2)\in[n_1M]^2}|\calB_{n_1,M}^i(t_1,t_2)|\\
&\leq (n_1M)^2\max_{(t_1,t_2)\in[n_1M]^2}|\calB_{n_1,M}^i(t_1,t_2)|\\
&\leq (n_1M^2)\max_{(t_1,t_2)\in[n_1M]^2}|\calB_{n_1,M}^{i,\rmc}(t_1,t_2)|\max_{(t_1,t_2)\in[n_1M]^2}\frac{|\calB_{n_1,M}^i(t_1,t_2)|}{|\calB_{n_1,M}^{i,\rmc}(t_1,t_2)|}.
\end{align}
Subsequent steps are algebra. Note that the key idea is to upper bound the size of $\calB_{n_1,M}^i$ with the size of $\calB_{n_1,M}^{i,\rmc}(t_1,t_2)$, which is more tractable. Using the same method, one can also lower bound $|\calB_{n_1,M}^i|$ by $\max_{(t_1,t_2)\in[n_1M]^2}|\calB_{n_1,M}^{i,\rmc}(t_1,t_2)|$. However, such a lower bound is too loose. It is of independent interest to derive tight bounds on $|\calB_{n_1,M}^i|$ and propose a low complexity method to generate the set $\calB_{n_1,M}$.

We then define the set of trajectories for other time slots. Fix any $j\in[2:B]$. Given initial location $\bs\in[0,1]^d$ and velocities of previous time slots $\bv^{j-1}\in\calV^{(j-1)\times d}$, the set of all possible quantized trajectories in the $j$-th time slot $[n_{j-1}+1,n_j]$ is
\begin{align}
\calB_{N_j,M}(\bs,\bv^{j-1})
&:=\Big\{\bw_{n_{j-1}+1}^{n_j}\in([N_jM])^{N_j\times d}:~\bw_{n_{j-1}+1}^{n_j}=\bw(\bs,\bv^j,[n_{j-1}+1,n_j])\mathrm{~for~some~}\bv_j\in\calV^d\Big\}\label{def:calBj},
\end{align}
where $\bw(\bs,\bv^j,[n_{j-1}+1,n_j]):=(w(\bs,\bv^j,n_{j-1}+1),\ldots,w(\bs,\bv^j,n_j))\in([N_jM])^{N_j\times d}$ is the trajectory of the target at the $j$-th time slot that moves constantly with velocity $\bv_j$ given $\bs$ and $\bv^{j-1}$. Similarly to \eqref{upptrajectories}, for any $\bs$ and $\bv^{j-1}$, we can upper bound the number of trajectories in the $j$-th time slot as
\begin{align}
|\calB_{N_j,M}(\bs,\bv^{j-1})|\leq ((2N_jv_++3)N_j^3M)^d\label{upptrajectoriesj}.
\end{align}
The trajectory sets $\calB_{n_1,M}$ and $\calB_{N_j,M}(\cdot)$ are used to present our proposed non-adaptive query procedure and also used in our proofs.

We will also need the following definitions of information densities. Given any $(p,q)\in[0,1]^2$, let $P_Y^{p,q}$ be the marginal distribution on $\calY$ that is induced by the Bernoulli distribution $P_X=\mathrm{Bern}(p)$ and the query-dependent channel $P_{Y|X}^{q}$. Define the following likelihood ratio which is known as the information density~\cite{polyanskiy2010finite,zhou2019twentyq}
\begin{align}
\imath_{p,q}(x;y)&:=\log\frac{P_{Y|X}^q(y|x)}{P_Y^{p,q}(y)},~\forall~(x,y)\in[0,1]\times\calY\label{def:MIdensity}.
\end{align}
Correspondingly, for any $(x^n,y^n)\in[0,1]^n\times\calY^n$, we define
\begin{align}
\imath_{p,q}(x^n;y^n)
&:=\sum_{t\in[n]}\imath_{p,q}(x_t;y_t)\label{def:empiricalMI}
\end{align}
as the empirical mutual information between i.i.d. sequences $x^n$ and $y^n$.

\subsection{Non-Asymptotic Bounds}
\label{sec:fbl}

\begin{algorithm}
\caption{Non-adaptive query procedure}
\label{procedure:nonadapt}
\begin{algorithmic}
\REQUIRE Velocity change time points $\bn=(n_1,\ldots,n_B)$, dimension $d$ and two design parameters $(M,p)$
\ENSURE Estimated trajectory $\hat{\bw}^{n_B}\in([0,1]^d)^{n_B}$ of a target that moves with unknown initial location $\bS\in[0,1]^d$ and piecewise constant velocities $\bV^B\in\calV^{d\times B}$
\\\hrulefill
\STATE \emph{Query and response for the first time slot:}
\STATE Partition the unit cube $[0,1]^d$ into $(n_1M)^d$ equal size disjoint cubes $\{\calS_{i_1,\ldots,i_d}\}_{(i_1,\ldots,i_d)\in[n_1M]^d}$
\STATE Generate $(n_1M)^d$ binary vectors $\{x^{n_1}(i_1,\ldots,i_d)\}_{(i_1,\ldots,i_d)\in[nM]^d}$ where each binary vector is generated i.i.d. according to the Bernoulli distribution with parameter $p$

\FOR {$t\in[1,n_1]$}
\STATE Pose the $t$-th query to the oracle that asks whether the target currently lies in $\calA_t$ where
\begin{align*}
\calA_t:=\bigcup_{(i_1,\ldots,i_d)\in[n_1M]^d:x_t(i_1,\ldots,i_d)=1}\calS_{i_1,\ldots,i_d}
\end{align*}
\STATE Obtain a noisy response $y_t$.
\ENDFOR 
\STATE \emph{Decoding for the first time slot:}
\STATE Collect noisy responses $y^{n_1}=(y_1,\ldots,y_{n_1})$
\STATE Generate an estimated trajectory $\hat{\bw}^{n_1}=(\hat{\bw}_1,\ldots,\hat{\bw}_{n_1})$ for the first stage using maximal information density decoding as
\begin{align}
\hat{\bw}^{n_1}
&=\argmax_{\bar{\bw}^{n_1}\in\calB_{n_1,M}}\imath_{p,f(p)}(x^{n_1}(\bar{\bw}^{n_1});y^{n_1}),
\end{align}
where $x^{n_1}(\bar{\bw}^{n_1})=(x_1(\bar{\bw}_1),\ldots,x_{n_1}(\bar{\bw}_{n_1}))$
\STATE Generate estimates $(\hat{\bs},\hat{\bv}_1)$ such that $\bw(\hat{\bs},\hat{\bw}_1,[n_1])=\hat{\bw}^{n_1}$

\FOR {$j\in[2:B]$}
\STATE \emph{Query and response for the $j$-th time slot:}
\STATE Set $N_j=n_j-n_{j-1}$
\STATE Partition the unit cube $[0,1]^d$ into $(N_jM)^d$ equal size disjoint cubes $\{\calS^j_{i_1,\ldots,i_d}\}_{(i_1,\ldots,i_d)\in[N_jM]^d}$
\STATE Generate $(N_jM)^d$ binary vectors $\{x^{n_1}(i_1,\ldots,i_d)\}_{(i_1,\ldots,i_d)\in[N_jM]^d}$ where each binary vector is generated i.i.d. according to the Bernoulli distribution with parameter $p$
\FOR {$t\in[n_{j-1}+1,n_j]$}
\STATE Pose the $t$-th query to the oracle that asks whether the target currently lies in $\calA_t$ where
\begin{align*}
\calA_t:=\bigcup_{(i_1,\ldots,i_d)\in[N_jM]^d:x_t(i_1,\ldots,i_d)=1}\calS_{i_1,\ldots,i_d}^j
\end{align*}
\STATE Obtain a noisy response $y_t$
\ENDFOR 
\STATE \emph{Decoding for the $j$-th time slot:}
\STATE Collect noisy responses $y^{n_j}_{n_{j-1}+1}=(y_{n_{j-1}+1},\ldots,y_{n_j})$
\STATE Generate an estimated trajectory $\hat{\bw}_{n_{j-1}+1}^{n_j}\in([0,1]^d)^{N_j}$ using $\hat{\bs}$ and $\hat{\bv}^{j-1}=(\hat{\bv}_1,\ldots,\hat{\bv}_{j-1})$ as
\begin{align}
\hat{\bw}_{n_{j-1}+1}^{n_j}
&=\argmax_{\bar{\bw}_{n_{j-1}+1}^{n_j}\in\calB_{N_j,M}(\hat{\bs},\hat{\bv}^{j-1})}\imath_{p,f(p)}(x^n(\bar{\bw}_{n_{j-1}+1}^{n_j});y_{n_{j-1}+1}^{n_j}).
\end{align}
\STATE Generate an estimate $\hat{\bv}_j$ such that $\bw(\hat{\bs},\hat{\bv}^j,[n_{j-1}+1,n_j])=\hat{\bw}_{n_{j-1}+1}^{n_j}$
\ENDFOR
\end{algorithmic}
\end{algorithm}

Recall that $\bn=(n_1,\ldots,n_B)\in\bbN^B$ are the ending time points of the sequence of $B$ time slots. We first provide a non-asymptotic achievability result. Given any $(n,k,p,v_+,\eta)\in\bbN^2\times(0,1)\times\bbR_+^2$, let 
\begin{align}
\zeta(n,k,p,v_+,\eta)
&:=2n\eta K c(f(p))-\lceil 2nv_+\rceil \min\{\log(p),\log(1-p)\}+d\log(2nv_++3)+kd\log n\label{def:zeta}.
\end{align}
\begin{theorem}
\label{ach:fbl}
Consider any discrete query-dependent channel satisfying \eqref{assump:continuouschannel}. For any $(M,p,v_+)\in\bbN^2\times(0,1)\times(0,0.5)$, there exists an $(\bn,d,\frac{B+1}{M},\varepsilon)$-non-adaptive query procedure such that
\begin{align}
\varepsilon
\nn&\leq \min\Big\{1,\exp(\zeta(n_1,4,p,v_+,\eta))M^{2d}\mathbb{E}_{P_{XY}^{n_1}}\big[\exp(-\imath_{p,f(p)}(X^{n_1};Y^{n_1}))\big]\Big\}+4n_1\exp(-2(n_1M)^d\eta^2)\\*
&\qquad+ \sum_{j\in[2:B]}\bigg(\min\Big\{1,\exp(\zeta(N_j,3,p,v_+,\eta))M^d\mathbb{E}_{P_{XY}^{N_j}}\big[\exp(-\imath_{p,f(p)}(X^{N_j};Y^{N_j}))\big]\Big\}+4N_j\exp(-2(N_jM)^d\eta^2)\bigg)\label{ach:fbl:bound},
\end{align}
where the distribution $P_{XY}$ is induced by the Bernoulli distribution $P_X$ with parameter $p$ and the query-dependent channel $P_{Y|X}^{f(p)}$, i.e.,
\begin{align}
P_{XY}(x,y)=P_X(x)P_{Y|X}^{f(p)}(y|x),~(x,y)\in\{0,1\}\times\calY\label{def:PXY}.
\end{align}
\end{theorem}
The proof of Theorem \ref{ach:fbl} is provided in Section \ref{proof:ach:fbl} using the non-adaptive query procedure in Algorithm \ref{procedure:nonadapt}. In Algorithm \ref{procedure:nonadapt}, the oracle is queried sequentially about the trajectories of the target in different time slots using random coding to generate each query and using maximal mutual information density decoding to generate estimates of trajectories. 

In the proof of Theorem \ref{ach:fbl}, we combine techniques used to prove the random coding union bound~\cite{polyanskiy2010finite}, the results concerning the number of trajectories in \cite[Section V]{kaspi2018searching} and the change of measure technique~\cite{csiszar2011information}. The first term in \eqref{ach:fbl:bound} upper bounds the excess-resolution probability for estimating the trajectory in the first time slot and each term inside the summation of $j\in[2:B]$ upper bounds the excess-resolution probability for estimating the trajectory in the $j$-th time slot. For each $j\in[2:B]$, the additional term $4n\exp(-2N_j^dM^d\eta^2)$ results from the stochastic nature of the random query matrix and upper bounds the probability of the rare event that a query region $\calA_t$ with size $|\calA_t|$ bounded away from $p$ is chosen for any $t\in[n_{j-1}+1,n_j]$. The multiplicative term $\exp(\zeta(N_j,3,p,v_+,\eta))$ captures the combined effect of the unknown velocity $\bV_j$ and the approximation error of the change of measure that is applied to replace the query-dependent channel $f_{Y|X}^{\calA_t}$ with the measurement-independent channel $P_{Y|X}^{f(p)}$ for each $t\in[n_{j-1}+1,n_j]$. The above illustration is also true for the case of $j=1$.

Unfortunately, Theorem \ref{ach:fbl} does not hold for a query-dependent AWGN channel that violates the continuity assumption in \eqref{assump:continuouschannel}. This problem can be easily fixed by adding one additional constraint on the noise power and then applying the change-of-measure technique to the truncated AWGN channel, as discussed in Section \ref{sec:mdAWGN}. This way, we obtain the following non-asymptotic bound for a query-dependent AWGN channel in Theorem \ref{ach:fbl:awgn}. Analogous to the definition of $\zeta(\cdot)$ in \eqref{def:zeta}, we need the following definition of $\tau(p,\eta,\alpha,\sigma)$ for any $(n,k,p,\eta,\sigma,v_+)\in\bbN^2\times(0,1)\times\calV\times\bbR_+^2$
\begin{align}
\tau(p,\eta,\alpha,\sigma)
&:=\frac{2(K\eta(f(p)+K\eta))\big(f(p)^2+4K\eta(f(p)+K\eta)\big)}{f(p)^2(f(p)^2-2K\eta(f(p)-K\eta))}\label{def:tau},\\
\zeta_\rmG(n,k,p,\eta,\sigma,v_+)
&:=\zeta(n,k,p,v_+,\eta)+n\tau(p,\eta,\alpha,\sigma)-2n\eta K c(f(p))\label{def:zeta:awgn}.
\end{align}
Our non-asymptotic achievability bound for a query-dependent AWGN channel is as follows.
\begin{theorem}
\label{ach:fbl:awgn}
For any $(M,p,\sigma,\eta,\alpha,\eta,v_+)\in\bbN\times(0,1)\times\bbR_+^4\times(0,0.5)$, there exists an $(\bn,d,\frac{B+1}{M},\varepsilon)$-non-adaptive query procedure such that
\begin{align}
\varepsilon
\nn&\leq  \min\Big\{1,\exp(\zeta_\rmG(n_1,4,p,\eta,\sigma,v_+))M^{2d}\mathbb{E}_{P_{XY}^{n_1}}\big[\exp(-\imath_{p,f(p)}(X^{n_1};Y^{n_1}))\big]\Big\}\\*
\nn&\qquad+\sum_{j\in[2:B]}\min\Big\{1,\exp(\zeta_\rmG(N_j,3,p,\eta,\sigma,v_+))M^d\mathbb{E}_{P_{XY}^{N_j}}\big[\exp(-\imath_{p,f(p)}(X^{N_j};Y^{N_j}))\big]\Big\}\\*
&\qquad+\sum_{j\in[B]}\bigg(4N_j\exp(-2N_j^dM^d\eta^2)+\exp\left(-\frac{N_j(1-\log 2)}{2}\right)\bigg)\label{var:awgn}.
\end{align}
\end{theorem}
The proof of Theorem \ref{ach:fbl:awgn} is similar to that of Theorem \ref{ach:fbl} and is provided in Section \ref{sec:mdAWGN}.  The remarks for Theorem \ref{ach:fbl} also apply here. The explanations of different terms are the same as in Theorem \ref{ach:fbl} except the additional term $\exp\left(-\frac{N_j(1-\log 2)}{2}\right)$ that quantifies the effect of truncating the channel output to satisfy the continuity assumption in \eqref{assump:continuouschannel}. 

Note that the performance in Theorem \ref{ach:fbl:awgn} is also achieved by the query procedure in Algorithm \ref{procedure:nonadapt}. Furthermore, when one considers a query-dependent BSC or AWGN channel, the maximal mutual information density decoding is equivalent to the following nearest neighbor decoding, i.e.,
\begin{itemize}
\item for the first time slot, 
\begin{align}
\hat{\bw}^{n_1}=\argmin_{\bar{\bw}^{n_1}\in\calB_{n_1,M}}\left\|x^{n_1}(\bar{\bw}^{n_1})-y^{n_1}\right\|^2\label{nnd1}.
\end{align}
\item for the $j$-th time slot with $j\in[2:B]$,
\begin{align}
\hat{\bw}_{n_{j-1}+1}^{n_j}=\argmin_{\hat{\bw}_{n_{j-1}+1}^{n_j}\in\calB_{N_j,M}(\hat{\bs},\hat{\bv}^{j-1})}\left\|x^{n_1}(\bar{\bw}^{n_1})-y^{n_1}\right\|^2.
\end{align}
\end{itemize}
Therefore, for both query-dependent BSC and AWGN channels, the decoding can be done \emph{universally} without knowing the statistics of the channel. The equivalence between maximal mutual information density decoding in Algorithm \ref{procedure:nonadapt} and nearest neighbor decoding is justified in Appendix \ref{just:nnd}.

Our next result provides a bound on the best performance of any non-adaptive query procedure. To present our result, given any $\calA\subseteq[0,1]^d$, using \eqref{def:MIdensity}, define the following query-dependent mutual information density
\begin{align}
\imath_{\calA,f}(X;Y)&:=\imath_{|\calA|,f(|\calA|)}(X;Y).
\end{align}
\begin{theorem}
\label{fbl:converse}
Consider any $(\delta,\varepsilon)\in\bbR_+\times(0,1)$. Given any $\beta\in(0,\frac{1-\varepsilon}{2})$ and any $\kappa\in(0,1-\varepsilon-2(1+4B)d\beta)$, any $(\bn,d,\delta,\varepsilon)$-non-adaptive query procedure satisfies
\begin{align}
-(B+1)d\log\delta
\nn&\leq\sup_{\calA^{n_B}\in([0,1]^d)^{n_B}}\sup\bigg\{r\Big|\Pr\Big\{\sum_{t\in[n_B]}\imath_{\calA_t,f}(X_t;Y_t)\leq r\Big\}\leq \varepsilon+2(1+4B)d\beta+\kappa\bigg\}\\
&\qquad-(B+1)d\log\beta-\sum_{j\in[B]}\log (2N_jv_+)-\log\kappa\label{conineq}.
\end{align}
\end{theorem}
The proof of Theorem \ref{fbl:converse} is provided in Section \ref{proof:fbl:converse}. Our proof proceeds in three steps. Firstly, we lower bound the excess-resolution probability of estimating the target trajectory by the excess-resolution probability of estimating the location $\bS$ and the velocities $\bV^B$. Subsequently, we show that the latter probability is further lower bounded by the error probability of transmitting a message over a query-dependent noisy channel with $n_B$ channel uses. Finally, we use the the non-asymptotic converse bound for channel coding~\cite[Proposition 4.4]{TanBook} to obtain the desired result. Note that our proof here is different from the converse arguments based on Fano's inequality in the proof \cite[Theorem 3]{kaspi2018searching}. 

We remark that the result in Theorem \ref{fbl:converse} holds for any query-dependent channel, regardless of the alphabet of the output. The bound in Theorem \ref{fbl:converse} is challenging to evaluate for relatively large $n_B$ due to the difficulty of optimizing over all possible choices of queries $\calA^{n_B}=(\calA_1,\ldots,\calA_{n_B})\in([0,1]^d)^{n_B}$. However, when $n_B$ is relatively large, the bound can be approximated accurately by much simpler expressions.

\subsection{Second-Order Asymptotics}
In this subsection, using the non-asymptotic bounds established in Section \ref{sec:fbl}, we derive the second-order approximation to the fundamental limit $\delta^*(n,d,\varepsilon)$. To present our results, we need the following definitions. Recall the definition of the mutual information density of $\imath_{p,f(p)}(\cdot)$ in \eqref{def:MIdensity}. For any query-dependent channel $\{P_{Y|X}^q\}_{q\in[0,1]}$, let
\begin{align}
C:=\max_{p\in(0,1)} \mathbb{E}[\imath_{p,f(p)}(X;Y)]\label{def:c},
\end{align}
where $(X,Y)\sim \mathrm{Bern}(p)\times P_{Y|X}^{f(p)}$. Let $\calP_{\rm{ca}}$ be the set of values $p$ achieving $C$. Furthermore, given any $p\in(0,1)$, let
\begin{align}
\rmV_p&:=\mathrm{Var}[\imath_{p,f(p)}(X;Y)]\label{def:vp}\\
\rmT_p&:=\mathbb{E}[|\imath_{p,f(p)}(X;Y)-\mathbb{E}[\imath_{p,f(p)}(X;Y)]|^3]\label{def:tp},
\end{align}
be the variance and the centered third absolute moment of the mutual information density $\imath_{p,f(p)}(X;Y)$. Finally, given any $\varepsilon\in(0,1)$, let
\begin{align}
\rmV_{\varepsilon}
&=\left\{
\begin{array}{ll}
\mathrm{max}_{p\in\calP_{\rm{ca}}}\rmV_p&\mathrm{if~}\varepsilon\leq 0.5,\\
\mathrm{min}_{p\in\calP_{\rm{ca}}}~\rmV_p&\mathrm{otherwise}.
\end{array}
\right.\label{def:v}
\end{align}

Recall that $\bn=(n_1,\ldots,n_B)$ are ending time points of $B$ different time slots of the piecewise velocity model in Section \ref{sec:piecewise}, $N_1=n_1$ and $N_j=n_j-n_{j-1}$ for each $j\in[2:B]$. Note that for any $p\in\calP_{\rm{ca}}$, the third absolute moment $\rmT_p$ is finite, i.e., $\rmT_p<\infty$, for any channel with discrete alphabet~\cite[Lemma 47]{polyanskiy2010finite} and any AWGN channel~\cite{tantomamichel2015}.
\begin{theorem}
\label{mainresult}
For any $\varepsilon\in(0,1)$ and any query-dependent channel satisfying \eqref{assump:continuouschannel}, given any $p\in\calP_{\mathrm{ca}}$, the minimal achievable resolution satisfies
\begin{align}
\nn&-d\log\delta^*(\bn,d,\varepsilon)\\*
\nn&\geq\max_{\substack{(\varepsilon_1,\ldots,\varepsilon_B):\\\sum_{j\in[B]}\varepsilon_j\leq \varepsilon}}\min\bigg\{\frac{n_1C+\sqrt{n_1\rmV_{\varepsilon_1}}\Phi^{-1}(\varepsilon_1)-\zeta(n_1,4,p,v_+,\eta)-\log n_1+O(1)}{2},\\*
&\qquad\qquad\qquad\qquad\min_{j\in[2:B]}\Big(N_jC+\sqrt{N_j\rmV_{\varepsilon_j}}\Phi^{-1}(\varepsilon_j)-\zeta(N_j,3,p,v_+,\eta)-\log N_j+O(1)\Big)\bigg\}-d\log(B+1)\label{sr:lower}.
\end{align}
The same is also true for a query-dependent AWGN channel with $\zeta(\cdot)$ replaced by $\zeta_\rmG(\cdot)$.
Conversely, for any $\varepsilon\in(0,1)$, the minimal achievable resolution satisfies
\begin{align}
-(B+1)d\log\delta^*(\bn,d,\varepsilon)
\leq n_BC+\sqrt{n_B\rmV_{\varepsilon}}\Phi^{-1}(\varepsilon)+O(\log n_B)\label{sr:upper}.
\end{align}
\end{theorem}
The proof of Theorem \ref{mainresult} is provided in Section \ref{proof:mainresult}, which follows by applying the Berry-Esseen theorem~\cite{berry1941accuracy,esseen1942liapounoff} to our derived non-asymptotic achievability and converse bounds in Theorems \ref{ach:fbl} to \ref{fbl:converse}. In the achievability proof, we only present details for a query-dependent discrete channel since the proof for a query-dependent AWGN channel is identical except that we need to replace $\zeta(\cdot)$ by $\zeta_\rmG(\cdot)$ and incorporate an additional term $\exp(-\frac{N_j(1-\log 2)}{2})$ into the definition of $\varepsilon_j$ in \eqref{def:varepsilonj}.

In general, the upper and lower bounds on $-\log\delta^*(\bn,d,\delta)$ in Eq. \eqref{sr:lower} and Eq. \eqref{sr:upper} do not match. We believe our achievability result is not tight since successive decoding is used slot-by-slot to estimate the trajectory of the target in Algorithm \ref{procedure:nonadapt}. This way, we have to constrain the sum of excess-resolution probabilities from each time slot to achieve a final excess-resolution probability $\varepsilon$. This is analogous to the fact that separate source and channel coding (SSCC) achieves worse performance as compared to joint source-channel coding, at least in second-order asymptotics~\cite{kostinajscc}, where the sum of error probabilities of source and channel coding are assumed to satisfy the final error probability constraint in SSCC. To close the gap, we would need to jointly decode the trajectory of the target for all time slots. However, the analyses of the excess-resolution probability for joint decoding appears quite challenging and is left as future work.

There are at least two cases where the bounds match: one is when $n_j=\frac{(j+1)n_B}{B+1}$ for all $j\in[B]$ and the other is when $B=1$. Note that the former case corresponds to using $\frac{2n_B}{B+1}$ queries in the first time slot to estimate two parameters $\bS$ and $\bV_1$ and using $\frac{n_B}{B+1}$ queries to estimate $\bV_j$ in the $j$-th time slot for each $j\in[2:B]$, while the latter case corresponds to searching for a moving target with unknown initial location $\bS$ and a constant moving velocity $\bV_1$.

Our results in Theorem \ref{mainresult} specialize to the first case as follows.
\begin{corollary}
\label{coro:first}
For each $j\in[B]$, let $n_j=\frac{(j+1)n_B}{B+1}$. When the maximal speed $v_+$ is such that $n_Bv_+=o(n_B)$, for any $(d,\varepsilon)\in\bbN\times(0,1)$, the asymptotic resolution decay rate satisfies
\begin{align}
\lim_{(n_B)\to\infty}\frac{-\log\delta^*(\bn,d,\varepsilon)}{n_B}=\frac{C}{(B+1)d}\label{strongconverse}.
\end{align}
\end{corollary}
The result in \eqref{strongconverse} is known as a strong converse result, which holds regardless of the excess-resolution probability $\varepsilon$. In other words, tolerating a larger excess-resolution probability cannot improve the asymptotic resolution decay rate.

We make several remarks on Corollary \ref{coro:first}. Firstly, our result in \eqref{strongconverse} generalizes the first order asymptotic result in \cite[Theorem 3]{kaspi2018searching}, which was constrained to $d=1$, $B=1$ and $\varepsilon\to 0$, to any finite $(d,B)\in\bbN^2$ and any excess-resolution probability $\varepsilon\in(0,1)$. Note that our results establish the asymptotic fundamental limit for optimal non-adaptive search for a moving target over a multidimensional region with piecewise constant velocity model, which is a better model for practical search problems where the target can move with variable velocity over a two or three dimensional region.

Secondly, one might wonder whether it is asymptotically optimal to search for the moving target with the piecewise constant velocity model using a cold restart search in each time slot, where for each $j\in[2:B]$, one estimates the trajectory in the $j$-th time slot and the velocity $\bV_j$ without using estimates $(\hat{\bS},\hat{\bV}^{j-1})$ obtained from the previous $j-1$ time slots. If one uses the query and response process process for the first time slot described in Algorithm \ref{procedure:nonadapt} for each $j\in[B]$, let $\delta_{\mathrm{sep}}^*(\bn,d,\delta)$ denote the achievable resolution. Following the same method used to prove the achievability part of Theorem \ref{mainresult}, for any $(d,\varepsilon)\in\bbN\times(0,1)$, it holds that under the same assumptions of Corollary \ref{coro:first},
\begin{align}
\liminf_{n_B\to\infty}\frac{-\log\delta_{\mathrm{sep}}^*(\bn,d,\varepsilon)}{n_B}\geq \frac{C}{2(B+1)d},
\end{align}
which is smaller than the asymptotic achievable resolution of our query procedure in Algorithm \ref{procedure:nonadapt} in Eq. \eqref{strongconverse}. Thus, asymptotically, one should avoid cold restart search and instead use estimates of initial location and velocities from previous time slots in order to achieve optimal performance.

We then state the second-order asymptotic result for $B=1$, which corresponds to searching for a moving target with constant velocity. In this case, we use $\delta^*(n,d,\varepsilon)$ to denote $\delta^*(\bn,d,\varepsilon)$ where $n_1:=n\in\bbN$ denotes the number of queries.
\begin{corollary}
\label{coro}
Fix $B=1$. For any $\varepsilon\in(0,1)$, the minimal achievable resolution $\delta^*(n,d,\varepsilon)$ satisfies
\begin{align}
-2d\log\delta^*(n,d,\varepsilon)
=
\left\{
\begin{array}{ll}
nC+O(nv_+)&\mathrm{if~}nv_+=O(n^\nu),~\nu\in[0.5,1]\\
nC+\sqrt{n\rmV_{\varepsilon}}\Phi^{-1}(\varepsilon)+O(nv_+)&\mathrm{if~}nv_+=O(n^\nu),~\nu\in(0,0.5),\\
nC+\sqrt{n\rmV_{\varepsilon}}\Phi^{-1}(\varepsilon)+O(\log n)&\mathrm{if~}nv_+=O(1).
\end{array}
\right.
\end{align}
\end{corollary}
We remark that the conditions in Corollary \ref{coro} on the maximal speed $v_+$ are chosen so that the achievability result in \eqref{sr:lower} matches the converse result in \eqref{sr:upper}. Our achievability result in \eqref{sr:lower} depends strongly on $v_+$ through the function $\zeta(\cdot)$ defined in \eqref{def:zeta} for the discrete noise and through the function $\zeta_\rmG(\cdot)$ defined in \eqref{def:zeta:awgn} for AWGN.

Corollary \ref{coro} generalizes \cite[Theorem 3]{kaspi2018searching} that only provided a first-order asymptotically tight result when $v_+\to 0$ and $n\to\infty$ for $d=1$. Specifically, Corollary \ref{coro} applies to arbitrary finite $d\in\bbN$ and identifies different regimes of $v_+$ as a non-increasing function of the number of queries $n$. Furthermore, it provides a stronger second-order asymptotically tight characterization for the minimal achievable resolution when $nv_+=o(\sqrt{n})$, i.e., the maximal speed $v_+$ satisfies $v_+=o(n^{-\frac{1}{2}})$.

\begin{figure}[tb]
\centering
\begin{tabular}{ccc}
\hspace{-.25in} \includegraphics[width=.45\columnwidth]{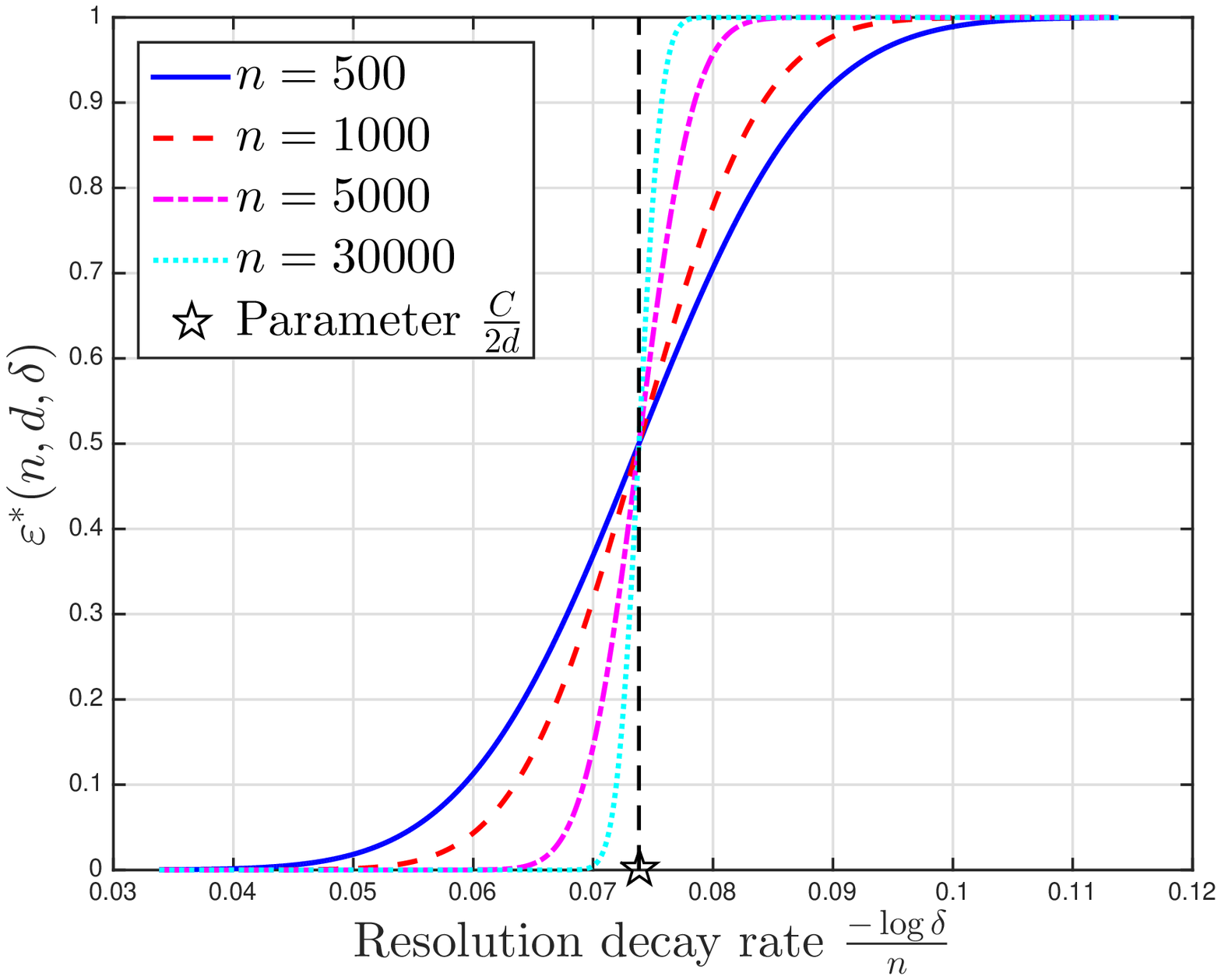}& \hspace{-.4in} \includegraphics[width=.45\columnwidth]{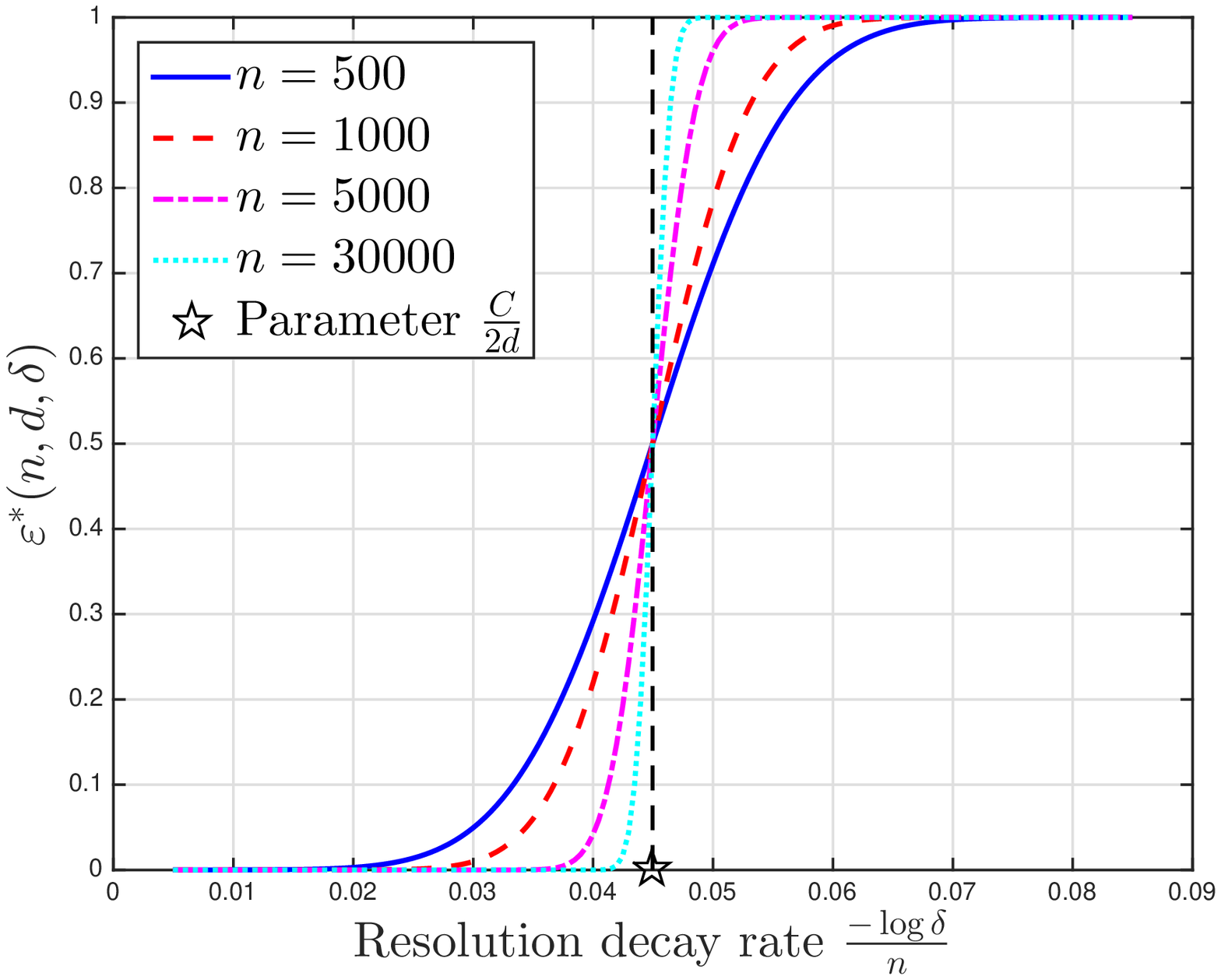}\\
\hspace{-.25in} {(a) query-dependent BSC with $\zeta=0.2$} & \hspace{-.4in}  { (b) query-dependent AWGN with $\sigma=2$}& \hspace{-.4in} \\
\end{tabular}
\caption{Illustration of the phase transition phenomenon of optimal non-adaptive query procedures for $d=1$. In both figures, the Lipschitz continuous size function $f(\cdot)$ in Definitions \ref{def:mdBSC} and \ref{def:mdAWGN} is chosen as $f(q)=2q+0.5$. Note that such a choice ensures that $\zeta f(q)\leq 0.5$ for all $q\in[0,1]$ so that the crossover probability for the query-dependent BSC is always no greater than half. The pentagram star denotes the critical phase transition threshold.}
\label{illus:phasetransition}
\end{figure}

Furthermore, Corollary \ref{coro} implies a phase transition phenomenon governing optimal non-adaptive query procedures when $B=1$. Specifically, if we let $\varepsilon^*(n,d,\delta)$ be the minimal excess-resolution probability of any non-adaptive query procedure with $n_1$ queries, when the maximal velocity $v_+$ is such that $nv_+=o(\sqrt{n})$, our result in Corollary \ref{coro} implies that
\begin{align}
\varepsilon^*(n,d,\delta)=\Phi\left(\frac{-d\log\delta-nC}{\sqrt{nV_\varepsilon}}\right)+o(1)\label{phaset}.
\end{align}
Eq. \eqref{phaset} implies the existence of a phase transition phenomenon for the minimal excess-resolution probability as a function of the resolution decay rate $-\frac{\log \delta}{n}$. In particular,  when the target resolution decay rate $\frac{-\log\delta}{n}$ is strictly greater than $\frac{C}{2d}$, the minimal excess-resolution probability tends to \emph{one} as the number of queries $n$ tends to infinity. On the other hand, when the target resolution decay rate is strictly less than the critical rate $\frac{C}{2d}$, the excess-resolution probability \emph{vanishes} as the number of queries $n$ increases. We numerically illustrate the phase transition phenomenon for a query-dependent BSC and an AWGN channel in Figure \ref{illus:phasetransition} for the case of $d=1$.

Comparing to \cite{zhou2019twentyq}, Corollary \ref{coro} asserts that under mild conditions, search for a $d$-dimensional moving target with unknown constant velocity over the unit cube of dimension $d$ has the same resolution as search for a $2d$-dimensional stationary target over the unit cube of dimension $2d$. However, as can be gleaned from the proofs of Corollary \ref{coro} and \cite[Theorem 4]{zhou2019twentyq}, the analyses for these two cases, especially the achievability part, are significantly different. This is because, when we search for a moving target with unknown velocity, the real time location of the target varies over time and is determined by both the initial location and the velocity. This makes the analysis for the moving target search problem complicated and requires consideration of intersected trajectories having different values of initial location and velocity, as demonstrated in Algorithm \ref{procedure:nonadapt}. In contrast, when we search for a stationary target, the location of the target is fixed over the search period.

\section{Proof of Main Results}
\label{sec:proofs}

\subsection{Proof of Theorem \ref{ach:fbl}}
\label{proof:ach:fbl}
\subsubsection{Preliminary definitions for target trajectories}
Given any $(n,M)\in\bbN^2$, partition the $d$-dimensional unit cube $[0,1]^d$ into $(nM)^d$ equal size disjoint regions $\{\calS_{i_1,\ldots,i_d}\}_{(i_1,\ldots,i_d)\in[nM]^d}$. We need the following additional definitions. For any $n\in\bbN$ and $\bv\in\calV^d$, define the following set of velocities
\begin{align}
\calD(\bv,n,M)
&:=\left\{\bar{\bv}\in\calV^d:~\max_{i\in[d]}|v_{j,i}-\barv_i|\leq\frac{1}{nM}\right\}\label{def:calDj},
\end{align}
Furthermore, for any $(\bs,\bv,n,M)\in[0,1]^d\times\calV^d\times\bbN^2$, define the following set of locations and velocities:
\begin{align}
\calD(\bs,\bv,n,M)
&:=\left\{(\bar{\bs},\bar{\bv})\in[0,1]^d\times\calD(\bv,n,M):\max_{i\in[d]}|s_i-\bars_i|\leq\frac{1}{M}\right\}\label{def:calD1}.
\end{align}
 
Recall the definitions of $l(\cdot)$ around Eq. \eqref{location}, $\bw(\cdot)$ around \eqref{quantizei}, $\calB_{n_1,M}$ in \eqref{def:calB1} and the description of the piecewise constant velocity model in Section \ref{sec:preliminaries} which uses $\bn=(n_1,\ldots,n_B)$ as the ending time points. Furthermore, recall that $N_j=n_j-n_{j-1}$ for each $j\in[B]$. Fix $\bs\in[0,1]^d$ and $\bv^B=(\bv_1,\ldots,\bv_B)\in\calV^{B\times d}$. For any $\bar{\bs}$ and $\bar{\bv}^B=(\bar{\bv}_1,\ldots,\bar{\bv}_B)\in(\calV^d)^B$ such that $(\bar{\bs},\bar{\bv}_1)\in\calD(\bs,\bv_1,n_1,M)$ and $\bar{\bv}_j\in\calD(\bv_j,N_j,M)$ for all $j\in[2:B]$, it follows that 
\begin{align}
\max_{j\in[B]}\max_{t\in[n_{j-1}+1,n_j]}\big\|l(\bar{\bs},\bar{\bv}^j,t)-l(\bs,\bv^j,t)\big\|_{\infty}\leq \frac{j+1}{M}\label{errorstep1}.
\end{align}
For subsequent analyses, we also need the following definitions of trajectories of the target in different time slots. For the first time slot, define the following set of trajectories
\begin{align}
\calU^{n_1}(\bs,\bv)
&:=\Big\{\bar{\bw}^{n_1}\in\calB_{n_1,M}:~\bar{\bw}^{n_1}=\bw(\bar{\bs},\bar{\bv},[n_1])~\mathrm{for~some~}(\bar{\bs},\bar{\bv})\notin\calD(\bs,\bv_1,n_1,M)\Big\}\label{def:calU1},
\end{align}
which can be equivalently written as the union of the following sets $\calU_l^{n_1}(\bs,\bv)$ over $l\in[n_1]$
\begin{align}
\calU_l^{n_1}(\bs,\bv)
&:=\Big\{\bar{\bw}^{n_1}\in\calU^{n_1}(\bs,\bv):~\sum_{t\in[n_1]}\bbo(\bar{\bw}_t=\bw(\bs,\bv,t))=l\Big\}\label{def:calU1l}.
\end{align} 
Note that the set $\calU^{n_1}(\bs,\bv)$ contains trajectories that are generated by all possible pairs of elements $(\bar{\bs},\bar{\bv})\in\calD(\bs,\bv_1,n_1,M)$ and and $\calU_l^{n_1}(\bs,\bv)$ contains the subset of trajectories that differ from $\bw(\bs,\bv,[n_1])$ for exactly $l$ times. In the following lemma, we present an upper bound on $l$, using \cite[Lemma 3]{kaspi2018searching}.
\begin{lemma}
\label{kaspi:lemma3}
Given any $(\bs,\bv)\in[0,1]^d\times\calV^d$ and $n_1\in\bbN$, for any $\tilde{\bw}^{n_1}=(\tilde{\bw}_1,\ldots,\tilde{\bw}_{n_1})\in\calU^{n_1}(\bs,\bv)$, the number of possible intersections of two trajectories $\bw(\bs,\bv,[n_1])$ and $\tilde{\bw}^{n_1}$ satisfies
\begin{align}
\big|\{t\in[n_1]:\bw(\bs,\bv,t)=\tilde{\bw}_t\}\big|
\leq \lceil2n_1v_+\rceil.
\end{align}
\end{lemma}
It follows from Lemma \ref{kaspi:lemma3} that
\begin{align}
\calU^{n_1}(\bs,\bv)=\bigcup_{l\in[2n_1v_+]}\calU_l^{n_1}(\bs,\bv)\label{unioncalUs}.
\end{align}
Furthermore, from the definitions of $\calB_{n_1,M}$ in \eqref{def:calB1} and $\calU^{n_1}(\bs,\bv)$ in \eqref{def:calU1}, it follows that
\begin{align}
|\calU^{n_1}(\bs,\bv)|\leq |\calB_{n_1,M}|\leq \big((2n_1v_++3)n_1^4M^2\big)^d\label{foruse}.
\end{align}
where \eqref{foruse} follows from \eqref{upptrajectories}.

Fix $j\in[2:B]$. Recall the definition of $\calB_{N_j,M}(\cdot)$ in \eqref{def:calBj}. Given $(\bs,\bv^j)$, define the sets of trajectories of the target in the $j$-th time slot as
\begin{align}
\nn&\calU^{N_j}(\bv_j|\bs,\bv^{j-1})\\*
&:=\Big\{\bar{\bw}_{n_{j-1}+1}^{n_j}\in\calB_{N_j,M}(\bs,\bv^{j-1}):~\bar{\bw}_{n_{j-1}+1}^{n_j}=\bw(\bs,(\bv^{j-1};\bar{\bv}_j),[n_{j-1}+1,n_j])~\mathrm{for~some~}\bar{\bv}_j\in\calD(\bv_j,N_j,M)\Big\}\label{def:calUj},
\end{align}
which can be equivalently written as the union of the following sets $\calU_l^{N_j}(\bv_j|\bs,\bv^{j-1})$ over $l\in[n_j-n_{j-1}]$, 
\begin{align}
\calU_l^{N_j}(\bv_j|\bs,\bv^{j-1})
:=\Big\{\bar{\bw}_{n_{j-1}+1}^{n_j}\in\calU^{N_j}(\bv_j|\bs,\bv^{j-1}):~\sum_{t\in[n_1]}\bbo(\bar{\bw}_t=\bw(\bs,\bv,t))=l\Big\}\label{def:calUjl}.
\end{align}
Similarly to \eqref{unioncalUs} and \eqref{foruse}, we have
\begin{align}
\calU^{N_j}(\bv_j|\bs,\bv^{j-1})&=\bigcup_{l\in[\lceil 2N_jv_+\rceil]}\calU_l^{N_j}(\bv_j|\bs,\bv^{j-1})\label{unioncalUj},\\
|\calU^{N_j}(\bv_j|\bs,\bv^{j-1})|&\leq |\calB_{N_j,M}(\bs,\bv^{j-1})|\leq \big((2N_jv_++3)N_j^3M\big)^d\label{foruse2}.
\end{align}
where \eqref{foruse2} follows form \eqref{upptrajectoriesj}.

\subsubsection{Query design and analysis}
\label{sec:search}
For any $(\bs,\bv^B)\in[0,1]^d\times\calV^{B\times d}$, our non-adaptive query procedure proceeds sequentially for different time slots as summarized in Algorithm \ref{procedure:nonadapt}. 

For the first time slot, we partition the unit cube $[0,1]^d$ into $(n_1M)^d$ equal size disjoint sub-cubes $\{\calS_{i_1,\ldots,i_d}\}_{(i_1,\ldots,i_d)\in[n_1M]^d}$\footnote{This can be done by equally partitioning each dimension into equal size sub-intervals and let $i_j$ be the index of the sub-interval of the cube in the $j$-th dimension for each $j\in[d]$.}. Furthermore, let $\bx^{n_1}=\{x^{n_1}(i_1,\ldots,i_d)\}_{(i_1,\ldots,i_d)\in[n_1M]}$ be the collection any $(n_1M)^d$ binary codewords that will be used as the query matrix. Given any possible trajectory $\bw^{n_1}=(\bw_1,\ldots,\bw_{n_1})\in[n_1M]^{n_1\times d}$, we use $x^{n_1}(\bw^{n_1})$ to denote the vector $(x_1(\bw_1),\ldots,x_n(\bw_n))$. The query and response procedure for the first time slot proceeds as follows. At each discrete time $t\in[n_1]$, one queries the oracle on whether the target is located in the region $\calA_t$ defined as follows:
\begin{align}
\calA_t=\bigcup_{(i_1,\ldots,i_d)\in[n_1M]^d:x_t(i_1,\ldots,i_d)=1}\calS_{i_1,\ldots,i_d}\label{def:query}.
\end{align}
Note that $\calA_t$ is the union of sub-cubes with index $(i_1,\ldots,i_d)$ such that the corresponding codeword $x_t(i_1,\ldots,i_d)$ is one. Given any $\bs\in[0,1]^d$ and moving velocity $\bv_1\in\calV^d$, the noiseless answer to the query $\calA_t$ is
\begin{align}
z_t
&=\bbo(\bw(\bs,\bv_1,t)\in\calA_t)
=\left\{
\begin{array}{cc}
1&\mathrm{if~}x_t(\bw(\bs,\bv,t))=1\\
0&\mathrm{otherwise}.
\end{array}
\right.
\end{align}
Since a query-dependent noisy channel is used to model the behavior of the oracle, the noisy answers $y_t$ is obtained by passing $z_t$ over the memoryless channel $P_{Y_t|Z_t}^{f(|\calA_t|)}(\cdot|z_t)$. Given noisy answers $y^{n_1}=(y_1,\ldots,y_{n_1})$, the decoder first generates estimates $\hat{\bw}^{n_1}=(\hat{\bw}_1,\ldots,\hat{\bw}_n)$ of the quantized trajectory using the following maximal mutual information density estimator:
\begin{align}
\hat{\bw}^{n_1}
&=\argmax_{\bar{\bw}^{n_1}\in\calB_{n_1,M}}\imath_{p,f(p)}(x^{n_1}(\bar{\bw}^{n_1});y^{n_1}),
\end{align}
where $\imath_{p,f(p)}(\cdot)$ is the empirical mutual information defined in \eqref{def:empiricalMI}. Subsequently, the decoder generates estimates $(\hat{\bs},\hat{\bv}_1)$ of $(\bs,\bv)$ as any $(\bar{\bs},\bar{\bv}_1)\in[0,1]^d\times\calV^d$ such that $\hat{\bw}^{n_1}=\bw(\bar{\bs},\bar{\bv}_1,[n_1])$.

We next describe the query and response procedure for the $j$-th time slot with $j\in[2:B]$. Recall that $N_j=n_j-n_{j-1}$. Similarly to the first time slot, we partition the unit cube $[0,1]^d$ into $(N_jM)^d$ equal size disjoint sub-cubes $\{\calS^j_{i_1,\ldots,i_d}\}_{(i_1,\ldots,i_d)\in[N_jM]^d}$. Furthermore, let $\bx_{n_{j-1}+1}^{n_j}=\{x_{n_{j-1}+1}^{n_j}(i_1,\ldots,i_d)\}_{(i_1,\ldots,i_d)\in[N_jM]}$ be the collection of $(N_jM)^d$ binary codewords. Given any possible trajectory $\bw_{n_{j-1}+1}^{n_j}=(\bw_{n_{j-1}+1},\ldots,\bw_{n_j})\in[n_1M]^{N_j\times d}$, we use $x_{n_{j-1}+1}^{n_j}(\bw_{n_{j-1}+1}^{n_j})$ to denote the vector $(x_{n_{j-1}+1}(\bw_{n_{j-1}+1}),\ldots,x_{n_j}(\bw_{n_j}))$. At each discrete time $t\in[n_{j-1}+1,n_j]$, the query asks whether the target current locates in the region $\calA_t$ defined in \eqref{def:query} with $\calS_{i_1,\ldots,i_d}$ replaced by $\calS^j_{i_1,\ldots,i_d}$. The noiseless response is $z_t=x_t(\bw(\bs,\bv^j,t))$ and the noisy answer $y_t$ is the output of passing $z_t$ over $P_{Y_t|Z_t}^{f(|\calA_t|)}(\cdot|z_t)$. Using noisy responses $y_{n_{j-1}+1}^{n_j}=(y_{n_{j-1}+1},\ldots,y_{n_j})$, the estimates $\hat{\bs}$ and $\hat{\bv}^{j-1}=(\hat{\bv}_1,\ldots,\hat{\bv}_{j-1})$, the decoder first generates estimations $\hat{\bw}_{n_{j-1}+1}^{n_j}=(\hat{\bw}_{n_{j-1}+1},\ldots,\hat{\bw}_{n_j})$ of the quantized trajectory as
\begin{align}
\hat{\bw}_{n_{j-1}+1}^{n_j}
&=\argmax_{\bar{\bw}_{n_{j-1}+1}^{n_j}\in\calB_{N_j,M}(\hat{\bs},\hat{\bv}^{j-1})}\imath_{p,f(p)}(x_{n_{j-1}+1}^{n_j}(\bar{\bw}_{n_{j-1}+1}^{n_j});y_{n_{j-1}+1}^{n_j}),
\end{align}
Subsequently, the decoder generates an estimate $\hat{\bv}_j\in\calV^d$ such that $\hat{\bw}_{n_{j-1}+1}^{n_j}=\bw(\bar{\bs},\bar{\bv}^j,[n_{j-1}+1,n_j])$. Finally, the estimated trajectory $\hat{\bw}^{n_B}$ is the concatenation of $\hat{\bw}_{n_{j-1}+1}^{n_j}$ for each $j\in[B]$.

Under the above non-adaptive query procedure, the conditional excess-resolution probability with respect to the resolution level $\frac{B+1}{M}$ given any  $(\bs,\bv^B)\in[0,1]^d\times\calV^{B\times d}$ satisfies
\begin{align}
\nn&\rmP_\rme\left(\bn,d,\frac{B+1}{M}\Big|\bs,\bv\right)\\*
&=\Pr\left\{\max_{t\in[0,n_1]}\|l(\hat{\bs},\hat{\bv}_1,t)-l(\bs,\bv_1,t)\|_{\infty}>\frac{B+1}{M}\mathrm{~or~}\max_{j\in[2:B]}\max_{t\in[n_{j-1}+1,n_j]}\|l(\hat{\bs},\hat{\bv}^j,t)-l(\bs,\bv^j,t)\|_{\infty}>\frac{B+1}{M}\right\}\\
\nn&=\Pr\bigg\{\|l(\hat{\bs},\hat{\bv}_1,t)-l(\bs,\bv_1,t)\|_{\infty}>\frac{B+1}{M}~\mathrm{for~some~}t\in[0,n_1],\\*
&\qquad\qquad\qquad\mathrm{or~}\exists~j\in[2:B]:~\|l(\hat{\bs},\hat{\bv}^j,t)-l(\bs,\bv^j,t)\|_{\infty}>\frac{B+1}{M}~\mathrm{for~some~}t\in[n_{j-1}+1,n_j]\bigg\}\\*
&\leq \Pr\Big\{(\hat{\bs},\hat{\bv}_1)\notin\calD(\bs,\bv_1,n_1,M)\mathrm{~or~}\exists~j\in[2:B]:~\hat{\bv}_j\notin\calD(\bv_j,N_j,M)\Big\}\label{useerro1}\\
&\leq \Pr\Big\{(\hat{\bs},\hat{\bv}_1)\notin\calD(\bs,\bv_1,n_1,M)\}+\sum_{j\in[2:B]}\Pr\{\hat{\bv}_j\notin\calD(\bv_j,N_j,M)\Big\}\\
\nn&\leq \Pr\Big\{\exists~\bw^{n_1}\in\calU^{n_1}(\bs,\bv_1):~\imath_{p,f(p)}(x^{n_1}(\tilde{\bw}^{n_1});Y^{n_1})\geq \imath_{p,f(p)}(x^{n_1}(\bw(\bs,\bv^j,[n_1]));Y^{n_1})\Big\}\\*
\nn&\qquad+\sum_{j\in[2:B]}\Pr\Big\{\exists~\tilde{\bw}_{n_{j-1}+1}^{n_j}\in\calU^{N_j}(\bv_j|s,\bv^{j-1}):~
\imath_{p,f(p)}(x_{n_{j-1}+1}^{n_j}(\tilde{\bw}_{n_{j-1}+1}^{n_j});Y_{n_{j-1}+1}^{n_j})\\*
&\qquad\qquad\qquad\qquad\qquad\qquad\qquad\qquad\qquad\qquad\qquad\geq \imath_{p,f(p)}(x_{n_{j-1}+1}^{n_j}(\bw(\bs,\bv^j,[n_{j-1}+1,n_j]));Y_{n_{j-1}+1}^{n_j})\Big\}\label{ach:move1},
\end{align}
where \eqref{useerro1} follows from the result in \eqref{errorstep1}. For ease of notation, we denote the first term in \eqref{ach:move1} as  $\rmP_\rme(\bx^{n_1},\bs,\bv_1)$ and the probability term for each $j\in[2:B]$ in the second term of \eqref{ach:move1} as $\rmP_\rme(\bx_{n_{j-1}+1}^{n_j},\bs,\bv^{j-1})$.

\subsubsection{Upper bound the excess-resolution probability for the first time slot}

To upper bound the probability terms, we employ a random coding argument. We first consider $\rmP_\rme(\bx^{n_1},\bs,\bv_1)$. To proceed, let $\bX^{n_1}=\{X^{n_1}(i_1,\ldots,i_d)\}_{(i_1,\ldots,i_d)\in[n_1M]^d}$ be the collection of $(n_1M)^d$ random binary vectors where each one is generated i.i.d. from a Bernoulli distribution $P_X$ with parameter $p$. The joint distribution $P_{\bX^{n_1}Y^{n_1}}^{(\mathrm{md})}$ of the query matrix $\bX^{n_1}$ and the noisy responses $Y^{n_1}$ for the first time slot satisfies that for any $(\bx^{n_1},y^{n_1})\in([0,1]^{n_1})^{n_1M}\times\calY^{n_1}$
\begin{align}
P_{\bX^{n_1} Y^{n_1}}^{(\rm{md},\bs,\bv_1)}(\bx^{n_1},y^{n_1})
&=\Big(\prod_{(i_1,\ldots,i_d)\in[n_1M]^d}P_X^{n_1}(x^{n_1}(i_1,\ldots,i_d))\Big)\prod_{t\in[n_1]}P_{Y|X}^{f(|\calA_t|)}(y_t|x_t(\bw(\bs,\bv_1,t))\label{def:pmd:move}.
\end{align}
To apply the change-of-measure technique, we also need the following alternative query-independent distribution
\begin{align}
P_{\bX^{n_1} Y^{n_1}}^{(\mathrm{alt},\bs,\bv_1)}(\bx^{n_1},y^{n_1})
&=\Big(\prod_{(i_1,\ldots,i_d)\in[n_1M]^d}P_X^{n_1}(x^{n_1}(i_1,\ldots,i_d))\Big)\prod_{t\in[n_1]}P_{Y|X}^{f(p)}(y_t|x_t(\bw(\bs,\bv_1,t))\label{def:alt:move}.
\end{align}
In the following analysis, the expectation operator $\mathbb{E}$ is calculated according to $P_{\bX,Y^{n_1}}^{(\mathrm{md},\bs,\bv)}$ or its induced distributions unless otherwise stated. Furthermore, for any $\eta\in\bbR_+$,
define the following typical set of typical query matrices 
\begin{align}
\calT^{n_1}(M,d,p,\eta)
&:=\Big\{\bx^{n_1}\in(\{0,1\}^{n_1})^{n_1M}:~\max_{t\in[n_1]}|q_{t,d}^{n_1,M}(\bx^{n_1})-p|\leq \eta\Big\}\label{def:calT1},
\end{align}
where we define
\begin{align}
q_{t,d}^{n_1,M}(\bx^{n_1})&:=\frac{1}{(n_1M)^d}\sum_{(i_1,\ldots,i_d)\in[n_1M]^d}x_t(i_1,\ldots,i_d)\label{def:tqd1}.
\end{align}
Similarly to the proof of \cite[Theorem 1]{zhou2019twentyq}, it follows that for any $\eta>0$,
\begin{align}
\mathbb{E}[\rmP_\rme(\bX^{n_1},\bs,\bv_1)]
&\leq \mathbb{E}[\rmP_\rme(\bX^{n_1},\bs,\bv_1)\bbo\{\bX^{n_1}\in\calT^{n_1}(M,d,p,\eta)\}]+\Pr\{\bX^{n_1}\notin\calT^{n_1}(M,d,p,\eta)\}\\
&\leq \mathbb{E}[\rmP_\rme(\bX^{n_1},\bs,\bv_1)\bbo\{\bX^{n_1}\in\calT^{n_1}(M,d,p,\eta)\}]+4n_1\exp(-2(n_1M)^d\eta^2)\label{ach:move2},
\end{align}
where the inequality follows from the upper bound of \cite[Lemma 22]{tan2014state}, which bounds the probability of atypicality of a random query matrix.

To further bound the result in \eqref{ach:move2}, let the joint distribution $P_{\barX XY}$ be defined such that for any $(\barx,x,y)\in[0,1]^2\times\calY]$,
\begin{align}
P_{\barX XY}(\barx,x,y)
&:=P_X(\barx)P_{XY}(x,y)=P_X(\barx)P_X(x)P_{Y|X}^p(y|x).
\end{align}
where the second equality follows from the definition of $P_{XY}$ in \eqref{def:PXY}. Applying techniques used to prove the random coding union bound in~\cite{polyanskiy2010finite}, we have
\begin{align}
\nn&\mathbb{E}[\rmP_\rme(\bX^{n_1},\bs,\bv_1)\bbo(\bX^{n_1}\in\calT^{n_1}(M,d,p,\eta))]\\*
&=\mathbb{E}\big[\bbo\{\bX^{n_1}\in\calT^{n_1}(M,d,p,\eta)\}\bbo\big\{\exists~\tilde{\bw}^{n_1}\in\calU^{n_1}(\bs,\bv)\mathrm{~:~}
\imath_{p,f(p)}(X^{n_1}(\tilde{\bw}^{n_1});Y^{n_1})\geq \imath_{p,f(p)}(X^{n_1}(\bw(\bs,\bv_1,[n_1]));Y^{n_1})\big\}\big]\\
&\leq\exp\big(2n_1\eta K c(f(p))\big) \mathbb{E}_{P_{\bX^{n_1} Y^{n_1}}^{(\mathrm{alt},\bs,\bv_1)}}\big[\bbo\big\{\exists~\tilde{\bw}^{n_1}\in\calU^{n_1}(\bs,\bv)\mathrm{~:~}
\imath_{p,f(p)}(X^{n_1}(\tilde{\bw}^{n_1});Y^{n_1})\geq \imath_{p,f(p)}(X^{n_1}(\bw(\bs,\bv_1,[n_1]));Y^{n_1})\big\}\big]\label{firstcomeasure}\\
\nn&=\exp\big(2n_1\eta K c(f(p))\big)\\*
&\qquad\times\sum_{l\in[\lceil 2n_1v_+\rceil]}\mathbb{E}_{P_{\bX^{n_1} Y^{n_1}}^{(\mathrm{alt},\bs,\bv_1)}}\big[\bbo\big\{\exists~\tilde{\bw}^{n_1}\in\calU_l^{n_1}(\bs,\bv)\mathrm{~:~}
\imath_{p,f(p)}(X^{n_1}(\tilde{\bw}^{n_1});Y^{n_1})\geq \imath_{p,f(p)}(X^{n_1}(\bw(\bs,\bv_1,[n_1]));Y^{n_1})\big\}\big]\label{usedef:calU}\\
&=\exp\big(2n_1\eta K c(f(p))\big) \sum_{l\in[\lceil 2n_1v_+\rceil]}|\calU_l^{n_1}(\bs,\bv)|\Pr_{P_{\barX XY}^{n_1}}\big\{\imath_{p,f(p)}(\barX^{n-l};Y^{n-l})\geq \imath_{p,f(p)}(X^{n-l};Y^{n-l})\big\}\label{usedefPbxxyk},\\
&\leq \exp\big(2n_1\eta K c(f(p))\big) \sum_{l\in[\lceil 2n_1v_+\rceil]}|\calU_l^{n_1}(\bs,\bv)|\mathbb{E}_{P_{XY}^{n-l}}\Big[\exp\big(-\imath_{p,f(p)}(X^{n-l};Y^{n-l})\big)\Big]\label{secondfomeasure}\\
&\leq \exp\big(2n_1\eta K c(f(p))\big) \sum_{l\in[\lceil 2n_1v_+\rceil]}|\calU_l^{n_1}(\bs,\bv)\exp(-l\min\{\log(p),\log(1-p)\}\mathbb{E}_{P_{XY}^{n_1}}[\exp(-\imath_{p,f(p)}(X^{n_1};Y^{n_1}))]\label{upimdensity}\\
&\leq \exp\Big(2n_1\eta K c(f(p))-\lceil 2n_1v_+\rceil \min\{\log(p),\log(1-p)\Big)|\calU^{n_1}(\bs,\bv)|\mathbb{E}_{P_{XY}^{n_1}}[\exp(-\imath_{p,f(p)}(X^{n_1};Y^{n_1}))]\label{usedefscalU}\\
&\leq \exp\Big(2n_1\eta K c(f(p))-\lceil 2n_1v_+\rceil \min\{\log(p),\log(1-p)\Big)(2n_1v_++3)^dn_1^{4d}M^{2d}\mathbb{E}_{P_{XY}^{n_1}}[\exp(-\imath_{p,f(p)}(X^{n_1};Y^{n_1}))]\label{ach:final}\\
&\leq \exp(\zeta(n_1,4,p,v_+,\eta))M^{2d}\mathbb{E}_{P_{XY}^{n_1}}[\exp(-\imath_{p,f(p)}(X^{n_1};Y^{n_1}))]\label{useeta}
\end{align}
where \eqref{firstcomeasure} follows from a change-of-measure, the fact that $f(\cdot)$ is Lipschitz continuous with parameter $K$ and the assumption on the query-dependent channel in \eqref{assump:continuouschannel}; \eqref{usedef:calU} and \eqref{usedefscalU} follow from the result in \eqref{unioncalUs} that specifies the relationship between $\calU^{n_1}(\bs,\bv_1)$ and $\calU^{n_1}_l(\bs,\bv_1)$; \eqref{usedefPbxxyk} follows since i) each codeword is generated i.i.d. from the same Bernoulli distribution and ii) the noisy channel is memoryless and time invariant under the joint distribution $P_{\bX^{n_1} Y^{n_1}}^{(\mathrm{alt},\bs,\bv_1)}$; \eqref{secondfomeasure} follows since for any real number $a\in\bbR$ and any $y^{n_1}\in\calY^{n_1}$, 
\begin{align}
P_X^{n_1}\{\imath_{p,f(p)}(\barX^{n_1};y^{n_1})\geq a\}
&=\sum_{\barx^{n_1}:\imath(\barx^{n_1};y^{n_1})\geq a}P_X^{n_1}(\barx^{n_1})\\
&\leq \sum_{\barx^{n_1}:P_X^{n_1}(\barx^{n_1})\leq \exp(-a)P_{X^{n_1}|Y^{n_1}}(\barx^{n_1}|y^{n_1})}\exp(-a)P_{X^{n_1}|Y^{n_1}}(\barx^{n_1}|y^{n_1})\\
&\leq \exp(-a);
\end{align}
\eqref{upimdensity} follows since for any $(x^{n_1},y^{n_1})\in[0,1]^{n_1}\times\calY^{n_1}$, the definition of the empirical mutual information $\imath_{p,q}(\cdot)$ in \eqref{def:empiricalMI} implies that
\begin{align}
\exp\big(-\imath_{p,f(p)}(x^{n-l};y^{n-l})\big)
&=\exp\bigg(-\imath_{p,f(p)}(x^{n_1};y^{n_1})+\sum_{j\in[n-l+1:n]}\imath_{p,f(p)}(x_j;y_j)\bigg)\\
&\leq \exp\bigg(-\imath_{p,f(p)}(x^{n_1};y^{n_1})+l\max_{(x,y)}\imath_{p,f(p)}(x;y)\bigg)\\
&\leq \exp\bigg(-\imath_{p,f(p)}(x^{n_1};y^{n_1})+l\max_{x\in\{0,1\}}\log\frac{1}{P_X(x)}\bigg)\\
&=\exp\bigg(-\imath_{p,f(p)}(x^{n_1};y^{n_1})-l\min\{\log(p),\log(1-p)\}\bigg),
\end{align}
\eqref{ach:final} follows from the result in \eqref{foruse}, and \eqref{useeta} follows the definition of $\zeta(\cdot)$ in \eqref{def:zeta}.

Since any probability term is no greater than one,
\begin{align}
\nn&\mathbb{E}[\rmP_\rme(\bX^{n_1},\bs,\bv_1)\bbo(\bX^{n_1}\in\calT^{n_1}(M,d,p,\eta))]\\*
&\leq \min\Big\{1,\exp(\zeta(n_1,4,p,v_+,\eta))M^{2d}\mathbb{E}_{P_{XY}^{n_1}}[\exp(-\imath_{p,f(p)}(X^{n_1};Y^{n_1}))]\Big\}.
\end{align}
Note that for any random variable $X$ and any constant $a$, $\mathbb{E}[X]\leq a$ implies that there exists $x\leq a$. Thus, there exists a deterministic query matrix  $\bx^{n_1}$ such that 
\begin{align}
\rmP_\rme(\bx^{n_1},\bs,\bv_1)
&\leq \min\Big\{1,\exp(\zeta(n_1,4,p,v_+,\eta))M^{2d}\mathbb{E}_{P_{XY}^{n_1}}[\exp(-\imath_{p,f(p)}(X^{n_1};Y^{n_1}))]\Big\}+4n_1\exp(-2(n_1M)^d\eta^2)\label{upp:excessp:slot1}.
\end{align}

\subsubsection{Final steps}
We next upper bound $\rmP_\rme(\bx_{n_{j-1}+1}^{n_j},\bs,\bv^{j-1})$. Similarly, we consider a random query matrix. Specifically, for each $j\in[2:B]$, let $\bX_{n_{j-1}+1}^{n_j}=\{X_{n_{j-1}+1}^{n_j}(i_1,\ldots,i_d)\}_{(i_1,\ldots,i_d)\in[N_jM]^d}\}$ be the collection of independent random binary vectors that are generated i.i.d. from $P_X$. Correspondingly, for each $j\in[B]$, let $Y_{n_{j-1}+1}^{n_j}$ be the noisy responses when this random query matrix is used. For each $j\in[2:B]$, given any $(\bx_{n_{j-1}+1}^{n_j},y_{n_{j-1}+1}^{n_j})\in([0,1]^{N_j})^{N_jM}\times\calY^{N_j}$, the joint distribution of the query matrix $\bX_{n_{j-1}+1}^{n_j}$ and the noisy responses $Y_{n_{j-1}+1}^{n_j}$ satisfies
\begin{align}
\nn&P_{\bX_{n_{j-1}+1}^{n_j}Y_{n_{j-1}+1}^{n_j}}^{(\mathrm{md},\bs,\bv^j)}(\bx_{n_{j-1}+1}^{n_j},y_{n_{j-1}+1}^{n_j})\\*
&=\Big(\prod_{(i_1,\ldots,i_d)\in[N_jM]^d}P_X^{N_j}(x_{n_{j-1}+1}^{n_j}(i_1,\ldots,i_d))\Big)\prod_{t\in[n_{j-1}+1,n_j]}P_{Y|X}^{f(|\calA_t|)}(y_t|x_t(\bw(\bs,\bv^j,t))\label{def:pmd:movej}.
\end{align}
We also need the following alternative query-independent distribution 
\begin{align}
\nn&P_{\bX_{n_{j-1}+1}^{n_j}Y_{n_{j-1}+1}^{n_j}}^{(\mathrm{alt},\bs,\bv^j)}(\bx_{n_{j-1}+1}^{n_j},y_{n_{j-1}+1}^{n_j})\\*
&=\Big(\prod_{(i_1,\ldots,i_d)\in[N_jM]^d}P_X^{N_j}(x_{n_{j-1}+1}^{n_j}(i_1,\ldots,i_d))\Big)\prod_{t\in[n_{j-1}+1,n_j]}P_{Y|X}^{f(p)}(y_t|x_t(\bw(\bs,\bv^j,t))\label{def:alt:movej},
\end{align}
Furthermore, for any $\eta\in\bbR_+$, define the following typical set of query matrices
\begin{align}
\calT^{N_j}(M,d,p,\eta)
&:=\Big\{\bx_{n_{j-1}+1}^{n_j}\in(\{0,1\}^{N_j})^{N_jM}:~\max_{t\in[n_{j-1}+1,n_j]}|q_{t,d}^{N_j,M}(\bx_{n_{j-1}+1}^{n_j})-p|\leq \eta\Big\},
\end{align}
where
\begin{align}
q_{t,d}^{N_j,M}(\bx_{n_{j-1}+1}^{n_j})&:=\frac{1}{(N_jM)^d}\sum_{(i_1,\ldots,i_d)\in[N_jM]^d}x_t(i_1,\ldots,i_d).
\end{align}
Similarly to \eqref{ach:move2}, for each $j\in[2:B]$ and any $\eta>0$, it follows that
\begin{align}
\mathbb{E}[\rmP_\rme(\bx_{n_{j-1}+1}^{n_j},\bs,\bv^{j-1})]
&\leq \mathbb{E}[\rmP_\rme(\bX_{n_{j-1}+1}^{n_j},\bs,\bv^{j-1})\bbo\{\bX_{n_{j-1}+1}^{n_j}\in\calT^{N_j}(M,d,p,\eta)\}]+\Pr\{\bX_{n_{j-1}+1}^{n_j}\notin\calT^{N_j}(M,d,p,\eta)\}\\
&\leq \mathbb{E}[\rmP_\rme(\bX_{n_{j-1}+1}^{n_j},\bs,\bv^{j-1})\bbo\{\bX_{n_{j-1}+1}^{n_j}\in\calT^{N_j}(M,d,p,\eta)\}]+4N_j\exp(-2(N_jM)^d\eta^2)\label{step1:achjerror}.
\end{align}
Using the same techniques as to derive \eqref{useeta}, we can upper bound the first term in \eqref{step1:achjerror} as follows:
\begin{align}
\nn&\mathbb{E}[\rmP_\rme(\bX_{n_{j-1}+1}^{n_j},\bs,\bv^{j-1})\bbo\{X_{n_{j-1}+1}^{n_j}\in\calT^{N_j}(M,d,p,\eta)\}]\\*
\nn&=\Pr\Big\{\bbo\{\bX_{n_{j-1}+1}^{n_j}\in\calT^{N_j}(M,d,p,\eta)\}\bbo\big\{\exists\tilde{\bw}_{n_{j-1}+1}^{n_j}\in\calU^{N_j}(\bv_j|s,\bv^{j-1}):\\*
&\qquad\qquad\qquad\qquad\imath_{p,f(p)}(X_{n_{j-1}+1}^{n_j}(\tilde{\bw}_{n_{j-1}+1}^{n_j});Y_{n_{j-1}+1}^{n_j})\geq \imath_{p,f(p)}(X_{n_{j-1}+1}^{n_j}(\bw(\bs,\bv^j,[n_{j-1}+1,n_j]));Y_{n_{j-1}+1}^{n_j})\big\}\Big\}\\
\nn&\leq \exp(2N_j\eta Kc(f(p)))\mathbb{E}_{P_{\bX_{n_{j-1}+1}^{n_j}Y_{n_{j-1}+1}^{n_j}}^{(\mathrm{alt},\bs,\bv^j)}}\Big[\bbo\big\{\exists\tilde{\bw}_{n_{j-1}+1}^{n_j}\in\calU^{N_j}(\bv_j|s,\bv^{j-1}):\imath_{p,f(p)}(X_{n_{j-1}+1}^{n_j}(\tilde{\bw}_{n_{j-1}+1}^{n_j});Y_{n_{j-1}+1}^{n_j})\\*
&\qquad\qquad\qquad\qquad\qquad\qquad\qquad\qquad\qquad\geq \imath_{p,f(p)}(X_{n_{j-1}+1}^{n_j}(\bw(\bs,\bv^j,[n_{j-1}+1,n_j]));Y_{n_{j-1}+1}^{n_j})\big\}\Big]\\
&\leq \exp(2N_j\eta Kc(f(p)))\sum_{l\in[\lceil 2N_jv_+\rceil]}\calU_l^{N_j}(\bv_j|\bs,\bv^{j-1})\mathbb{E}_{P_{XY}^{N_j-l}}\Big[\exp\big(-\imath_{p,f(p)}(X^{N_j-l};Y^{N_j-l})\big)\Big]\label{usecalUj}\\
&\leq \exp(\zeta(N_j,3,p,v_+,\eta))M^d\mathbb{E}_{P_{XY}^{N_j}}\big[\exp(-\imath_{p,f(p)}(X^{N_j};Y^{N_j}))\big]\label{usezetaandothers}.
\end{align}
where \eqref{usecalUj} follows form the union bound and the result in \eqref{unioncalUj} and \eqref{usezetaandothers} follows from the definition of $\zeta(\cdot)$ in \eqref{def:zeta} and the upper bound on the size of $\calU^{N_j}(\bv_j|s,\bv^{j-1})$ in \eqref{foruse2}. Since any probability is less than one, we further have
\begin{align}
\nn&\mathbb{E}[\rmP_\rme(\bX_{n_{j-1}+1}^{n_j},\bs,\bv^{j-1})\bbo\{X_{n_{j-1}+1}^{n_j}\in\calT^{N_j}(M,d,p,\eta)\}]\\*
&\leq\min\Big\{1,\exp(\zeta(N_j,3,p,v_+,\eta))M^d\mathbb{E}_{P_{XY}^{N_j}}\big[\exp(-\imath_{p,f(p)}(X^{N_j};Y^{N_j}))\big]\Big\}\label{lessthan1}.
\end{align}

Note that \eqref{lessthan1} implies that there exists a deterministic query matrix  $\bx_{n_{j-1}+1}^{n_j}$ such that 
\begin{align}
\rmP_\rme(\bx_{n_{j-1}+1}^{n_j},\bs,\bv^{j-1})
&\leq \Big\{1,\exp(\zeta(N_j,3,p,v_+,\eta))M^d\mathbb{E}_{P_{XY}^{N_j}}\big[\exp(-\imath_{p,f(p)}(X^{N_j};Y^{N_j}))\big]\Big\}+4N_j\exp(-2(N_jM)^d\eta^2)\label{upp:excessp:slotj}.
\end{align}

Combining \eqref{ach:move1}, \eqref{upp:excessp:slot1} and \eqref{upp:excessp:slotj} and noting that $N_1=n_1$ by definition, we conclude that the conditional excess-resolution probability with respect to the resolution level $\frac{B+1}{M}$ given any  $(\bs,\bv^B)\in[0,1]^d\times\calV^{B\times d}$ satisfies
\begin{align}
\nn&\rmP_\rme\left(\bn,d,\frac{B+1}{M}\Big|\bs,\bv\right)\\*
\nn&\leq \min\Big\{1,\exp(\zeta(n_1,4,p,v_+,\eta))M^{2d}\mathbb{E}_{P_{XY}^{n_1}}\big[\exp(-\imath_{p,f(p)}(X^{n_1};Y^{n_1}))\big]\Big\}+4n_1\exp(-2(n_1M)^d\eta^2)\\*
&\qquad+ \sum_{j\in[2:B]}\bigg(\min\Big\{1,\exp(\zeta(N_j,3,p,v_+,\eta))M^d\mathbb{E}_{P_{XY}^{N_j}}\big[\exp(-\imath_{p,f(p)}(X^{N_j};Y^{N_j}))\big]\Big\}+4N_j\exp(-2(N_jM)^d\eta^2)\bigg)\label{ach:complete}.
\end{align}
The proof of Theorem \ref{ach:fbl} is completed by noting that \eqref{ach:complete} holds uniformly for any $(\bs,\bv^B)$ and the right hand side of \eqref{ach:complete} upper bounds the excess-resolution probability under any joint distribution $f_{\bS\bV^B}$ of initial location $\bS$ and moving velocities $\bV^B$.

\subsection{Proof of Theorem \ref{ach:fbl:awgn}}
\label{sec:mdAWGN}
In this subsection, we emphasize the changes needed to prove Theorem \ref{ach:fbl:awgn} based on the proof of Theorem \ref{ach:fbl} since the same query procedure is used.

Fix $p\in(0,1)$. Let $Z^{n_1}$ be generated i.i.d. generated from the Gaussian distribution $\calN(0,\sigma^2)$ with mean $0$ and variance $\sigma^2$. Define $\alpha$ as
\begin{align}
\alpha=2\sigma^2\big(f(p)^2+4K\eta(f(p)+K\eta)\big),
\end{align}
Then, for any $(\bs,\bv_1)\in[0,1]^d\times\calV^d$ and any $\bx^{n_1}\in\calT^{n_1}(M,d,\eta,p)$, given queries $\calA^{n_1}=(\calA_1,\ldots,\calA_{n_1})$, we have
\begin{align}
\Pr\{\|Y^{n_1}-x^{n_1}(\bw(\bs,\bv_1,[n_1]))\|^2>n\alpha\}
&=\Pr\bigg\{\sum_{t\in[n_1]}f(|\calA_t|)^2Z_t^2>n\alpha\bigg\}\\
&\leq \Pr\bigg\{\frac{1}{n_1}\sum_{t\in[n_1]}Z_t^2>\frac{\alpha}{f(p)^2+4K\eta(f(p)+K\eta)}\bigg\}\label{usepropf}\\
&=\Pr\bigg\{\frac{1}{n_1}\sum_{t\in[n_1]}Z_t^2>2\sigma^2\bigg\}\\
&\leq \exp\left(-\frac{n_1(1-\log 2)}{2}\right)\label{prob_atyp_noise},
\end{align}
where \eqref{usepropf} follows since when $\bx^{n_1}\in\calT^{n_1}(M,d,\eta,p)$, the query size $|\calA_t|\in[p-\eta,p+\eta]$ for each $t\in[n_1]$ and thus
\begin{align}
f(|\calA_t|)^2
&\leq \max_{q\in[p-\eta,p+\eta]} f(q)^2\\
&=\max_{q\in[p-\eta,p+\eta]}(f(p)^2+f(q)^2-f(p)^2)\\
&\leq \max_{q\in[p-\eta,p+\eta]}\Big(f(p)^2+|f(q)-f(p)|\big(2f(p)+|f(q)-f(p)|\big)\Big)\\
&\leq \max_{q\in[p-\eta,p+\eta]}\Big(f(p)^2+K|q-p|\big(2f(p)+K|q-p|\big)\Big)\\
&\leq f(p)^2+2K\eta(f(p)+K\eta)\label{uppf},
\end{align}
and \eqref{prob_atyp_noise} follows from the Chernoff bound~\cite[Theorem B.4.1]{bouleau1994numerical} and the fact that $\mathbb{E}[Z_i]^2=\sigma^2$ for each $i\in[n]$, the justification of which is provided in Appendix \ref{justify:prob_noise}.

Recall the definition of $\tau(p,\eta,\alpha,\sigma)$ in \eqref{def:tau}. The analysis for the case of query-dependent AWGN case is exactly the same to that of Theorem \ref{ach:fbl} till \eqref{def:tqd1}. Then we need to bound the first term in \eqref{ach:move2} in a slightly different manner as follows:
\begin{align}
\nn&\mathbb{E}[\rmP_\rme(\bX^{n_1},\bs,\bv_1)\bbo\{\bX^{n_1}\in\calT^{n_1}(M,d,p,\eta)\}]\\*
\nn&\leq\mathbb{E}\Big[\rmP_\rme(\bX^{n_1},\bs,\bv_1)\bbo\{\bX^{n_1}\in\calT^{n_1}(M,d,p,\eta)\}\bbo(\|Y^{n_1}-X^{n_1}(\bw(\bs,\bv_1,[n_1]))\|^2\leq n\alpha)\Big]\\*
&\qquad+\mathbb{E}[\bbo\{\bX^{n_1}\in\calT^{n_1}(M,d,p,\eta)\}\bbo(\|Y^{n_1}-X^{n_1}(\bw(\bs,\bv_1,[n_1]))\|^2>n\alpha)]\\
&\leq \mathbb{E}\Big[\rmP_\rme(\bX^{n_1},\bs,\bv_1)\bbo\{\bX^{n_1}\in\calT^{n_1}(M,d,p,\eta)\}\bbo(\|Y^{n_1}-X^{n_1}(\bw(\bs,\bv_1,[n_1]))\|^2\leq n\alpha)\Big]+\exp\left(-\frac{n_1(1-\log 2)}{2}\right)\label{useprob},
\end{align}
where \eqref{useprob} follows from the result in \eqref{prob_atyp_noise}. Subsequently, the first term in \eqref{useprob} is further upper bounded as follows: 
\begin{align}
\nn&\mathbb{E}\Big[\rmP_\rme(\bX^{n_1},\bs,\bv_1)\bbo\{\bX^{n_1}\in\calT^{n_1}(M,d,p,\eta)\}\bbo(\|Y^{n_1}-X^{n_1}(\bw(\bs,\bv_1,[n_1]))\|^2>n\alpha)\Big]\\*
\nn&=\mathbb{E}\big[\bbo\{\bX^{n_1}\in\calT^{n_1}(M,d,p,\eta)\}\bbo\{\bX^{n_1}\in\calT^{n_1}(M,d,p,\eta)\}\bbo(\|Y^{n_1}-X^{n_1}(\bw(\bs,\bv_1,[n_1]))\|^2>n\alpha)\\*
&\qquad\qquad\times \bbo\big\{\exists~\tilde{\bw}^{n_1}\in\calU^{n_1}(\bs,\bv)\mathrm{~:~}
\imath_{p,f(p)}(X^{n_1}(\tilde{\bw}^{n_1});Y^{n_1})\geq \imath_{p,f(p)}(X^{n_1}(\bw(\bs,\bv_1,[n_1]));Y^{n_1})\big\}\big]\\
\nn&\leq \exp(n\tau(p,\eta,\alpha,\sigma))\mathbb{E}_{P_{\bX^{n_1} Y^{n_1}}^{(\mathrm{alt},\bs,\bv_1)}}\big[\bbo\big\{\exists~\tilde{\bw}^{n_1}\in\calU^{n_1}(\bs,\bv)\mathrm{~:~}
\imath_{p,f(p)}(X^{n_1}(\tilde{\bw}^{n_1});Y^{n_1})\\*
&\qquad\qquad\qquad\qquad\qquad\qquad\qquad\qquad\qquad\qquad\qquad\qquad\qquad\geq \imath_{p,f(p)}(X^{n_1}(\bw(\bs,\bv_1,[n_1]));Y^{n_1})\big\}\big]\label{cofmeasure4AWGN},
\end{align}
where \eqref{cofmeasure4AWGN} follows from a change-of-measure similarly to \eqref{firstcomeasure}. Following exactly the same steps till \eqref{useeta}, we can prove the result in \eqref{upp:excessp:slot1} with $\zeta(\cdot)$ replaced by $\zeta_\rmG(\cdot)$.

Similarly to steps leading to \eqref{cofmeasure4AWGN}, we can obtain results in \eqref{upp:excessp:slotj} with  $\zeta(\cdot)$ replaced by $\zeta_\rmG(\cdot)$ for each $j\in[2:B]$. The proof of Theorem \ref{ach:fbl:awgn} is then completed in the same manner as that of Theorem \ref{ach:fbl}.

\subsection{Proof of Theorem \ref{fbl:converse} }
\label{proof:fbl:converse}

We now prove the non-asymptotic converse result in Theorem \ref{fbl:converse} that bounds the performance of an optimal non-adaptive query procedure. Consider any sequence of queries $\calA^{n_B}=(\calA_1,\ldots,\calA_{n_B})\subseteq([0,1]^d)^{n_B}$ and any decoding function $g:\calY^{n_B}\to[0,1]^d\times\calV^{B\times d}$ such that the excess-resolution probability with respect to $\delta\in\bbR_+$ is upper bounded by $\varepsilon\in(0,1)$, i.e.
\begin{align}
\nn&\rmP_\rme(\bn,d,\delta)\\*
&=\sup_{f_{\bS\bV^B}}\Pr\Big\{\max_{t\in[0,n_1]}\|l(\hat{\bS},\hat{\bV}_1,t)-l(\bS,\bV_1,t)\|_{\infty}>\delta\mathrm{~or~}\max_{j\in[2:B]}\max_{t\in[n_{j-1}+1,n_j]}\|l(\hat{\bS},\hat{\bV}^j,t)-l(\bS,\bV^j,t)\|_{\infty}>\delta\Big\}\leq \varepsilon\label{error4converse},
\end{align}
where $\hat{\bS}$ and $\hat{\bV}^B=(\hat{\bV}_1,\ldots,\hat{\bV}_B)$ are estimates of the initial location $\bS$ and moving velocities $\bV^B$, respectively. Since the inequality in \eqref{error4converse} holds for arbitrary joint pdf $f_{\bS\bV^B}$. Without loss of generality, in subsequent analyses, we consider the joint pdf $f_{\bS\bV^B}^{\mathrm{unif}}$ such that random variables $(\bS,\bV^B)$ are independent of each other, $\bS$ distributes uniformly over $[0,1]^d$ and $\bV_j$ distributes uniformly over $\calV$ for each $j\in[B]$.

\subsubsection{Connection to the excess-resolution probability of estimating the location and velocities}

In this subsection, we show that the excess-resolution probability of estimating the trajectory in \eqref{error4converse} can be further lower bounded by the excess-resolution probability of estimating the initial location and velocities. To do so, we need to define the following events:
\begin{align}
\calE_0&:=\{(\hat{\bS},\hat{\bV}^B):~\|\hat{\bS}-\bS\|_{\infty}>\delta\},\\
\calE_1&:=\big\{(\hat{\bS},\hat{\bV}^B):~\|\hat{\bS}-\bS\|_{\infty}\leq \delta,~N_1\|\hat{\bV}_1-\bV_1\|_{\infty}>2\delta\big\},~\\
\calE_j&:=\bigg\{(\hat{\bS},\hat{\bV}^B):~\|\hat{\bS}-\bS\|_{\infty}\leq \delta,~\max_{l\in[j-1]}N_l\|\hat{\bV}_l-\bV_l\|_{\infty}\leq2\delta,~N_j\|\hat{\bV}_j-\bV\|_{\infty}>2\delta\bigg\},~j\in[2:B].
\end{align}
Note that $\calE_j$ are mutually independent for each $j\in[0:B]$. In the following, we show that the union of events $\bigcup_{j\in[0:B]}\calE_j$ leads to excess-resolution events in the probability term of \eqref{error4converse}. This is done inductively. Note that $\calE_0$ leads to an excess-resolution event at time $t=0$. Recall that we define $N_1=n_1$. Using the triangle inequality, it follows that if $(\hat{\bS},\hat{\bV}^B)\in\calE_1$,
\begin{align}
\|l(\hat{\bS},\hat{\bV}_1,n_1)-l(\bS,\bV_1,n_1)\|_{\infty}
&=\max_{i\in[d]}|(\hatS_i+n_1\hat{V}_{1,i})-(S_i+n_1V_{1,i})|\\
&\geq \max_{i\in[d]}\Big(n_1|\hatV_{1,i}-V_{1,i}|-|\hatS_i-S_i|\Big)\\
&>\frac{2n_1\delta}{N_1}-\delta\\
&=\delta,
\end{align}
which leads to an excess-resolution event at time $t=n_1$. For subsequent analyses, we need to further partition each set $\calE_j$ with $j\in[2:B]$ into the following two independent subsets:
\begin{align}
\calE_j^1&:=\{(\hat{\bS},\hat{\bV}^B)\in\calE_j:~\|l(\hat{\bS},\hat{\bV}^{j-1},n_{j-1})-l(\bS,\bV^{j-1},n_{j-1})\|_{\infty}>\delta\},\\
\calE_j^2&:=\{(\hat{\bS},\hat{\bV}^B)\in\calE_j:~\|l(\hat{\bS},\hat{\bV}^{j-1},n_{j-1})-l(\bS,\bV^{j-1},n_{j-1})\|_{\infty}\leq\delta\}.
\end{align}
The error event $\calE_j^1$ directly leads to an excess-resolution event at time $t=n_{j-1}$. If $(\hat{\bS},\hat{\bV}^B)\in\calE_j^2$,
\begin{align}
\|l(\hat{\bS},\hat{\bV}^j,n_j)-l(\bS,\bV^j,n_j)\|_{\infty}
&=\max_{i\in[d]}\big|(l(\hat{\bS},\hat{\bV}^{j-1},n_{j-1})+N_j\hatV_{j,i})-(l(\bS,\bV^{j-1},n_{j-1})+N_jV_{j,i})\big|\\
&\geq \max_{i\in[d]}\Big(N_j|\hatV_{j,i}-V_{j,i}|-\big|l(\hat{\bS},\hat{\bV}^{j-1},n_{j-1})-l(\bS,\bV^{j-1},n_{j-1})\big|\Big)\\
&>2\delta-\delta=\delta,
\end{align}
which leads to an excess-resolution event at time $t=n_j$.

Combining the above arguments and using \eqref{error4converse}, it follows that
\begin{align}
\varepsilon
&\geq \Pr\Big\{\max_{t\in[0,n_1]}\|l(\hat{\bS},\hat{\bV}_1,t)-l(\bS,\bV_1,t)\|_{\infty}>\delta\mathrm{~or~}\max_{j\in[2:B]}\max_{t\in[n_{j-1}+1,n_j]}\|l(\hat{\bS},\hat{\bV}^j,t)-l(\bS,\bV^j,t)\|_{\infty}>\delta\Big\}\\
&\geq \Pr\Big\{(\hat{\bS},\hat{\bV}^B)\in\bigcup_{j\in[0:B]}\calE_j\Big\}\\
&=\Pr\Big\{\|\hat{\bS}-\bS\|_{\infty}>\delta\mathrm{~or~}\exists~j\in[B]:~\|\hat{\bV}_j-\bV_j\|>\frac{2\delta}{N_j}\Big\}\label{justerror}.
\end{align}

\subsubsection{Connection to data transmission over a query-dependent noisy channel}
In this subsection, we further lower bound the right hand side of \eqref{justerror} by the probability of incorrectly decoding messages over a query-dependent noisy channel. To do so, we need to define quantization functions for the initial location and velocities. Let $\beta\in\bbR_+$ be arbitrary such that $\beta\leq\frac{1-\varepsilon}{2}<0.25$ and let $\barM:=\lceil\frac{\beta}{\delta}\rceil$. Partition the set $[0,1]$ into $\barM$ disjoint equal size subsets $\calS_1,\ldots,\calS_{\barM}$. Furthermore, for each $j\in[B]$, partition the set $[-v_+,v_+]$ into $\barM_j=\lceil N_jv_+\barM\rceil$ disjoint equal size subsets $\calV_1^j,\ldots,\calV_{\barM_j}^j$. Now define the following quantization functions
\begin{align}
\rmq_\rms(s)&:=\sum_{i\in[\barM]}i\bbo(s\in\calS_i),~\forall~s\in[0,1],\label{quan:sconverse}\\
\rmq_\rmv^j(v)&:=\sum_{i\in[\barM_j]}i\bbo(v\in\calV_i^j),~\forall~v\in[-v_+,v_+].
\end{align}

Given any sequence of queries $\calA^{n_B}=(\calA_1,\ldots,\calA_{n_B})\in([0,1]^d)^{n_B}$, the noiseless response $X^{n_B}=(X_1,\ldots,X_{n_B})$ is a sequence of independent random variables where for each $t\in[n_B]$, $X_t$ is a Bernoulli random variable with parameter being the volume $|\calA_t|$ of the query $\calA_t$. The noisy responses $Y^{n_B}=(Y_1,\ldots,Y_{n_B})$ are the output of passing $X^{n_B}$ into the query-dependent memoryless channel $\prod_{t\in[n_B]}P_{Y_t|X_t}^{\calA_t}$. Given $Y^{n_B}$, the decoder $g$ outputs estimate $(\hat{\bS},\hat{\bV}^B)=g(Y^{n_1})$ as the estimates for the initial location $\bS\in[0,1]^d$ and velocities $\bV^B\in\calV^{B\times d}$ of the moving target. In subsequent analyses, for each $i\in[d]$ and $j\in[B]$, let $W_{0,i}:=\rmq_\rms(S_i)$, $W_{j,i}=\rmq_\rmv^j(V_{j,i})$, $\hatW_{0,i}:=\rmq_\rms(\hatS_i)$ and $\hatW_{j,i}=\rmq_\rmv(\hatV_{j,i})$. Furthermore, let $\bW_0=(W_{0,i},\ldots,W_{0,d})$, $\bW_j=(W_{j,1},\ldots,W_{j,d})$ and we use $\hat{\bW}_0$ and $\hat{\bW}_j$ similarly. From the problem formulation, the Markov chain $(\bW_0,\bW_1,\ldots,\bW_j)-(\bS,\bV^B)-X^{n_B}-Y^{n_B}-(\hat{\bS},\hat{\bV}^B)$ holds and the joint distribution of these random variables satisfy
\begin{align}
\nn&P_{\bW_{[0:j]}\bS\bV^B X^{n_B}Y^{n_B}\hat{\bS}\hat{\bV}^B}(\bw_{[0:j]},\bs,\bv^B,x^{n_B},y^{n_B},\hat{\bs},\hat{\bv}^B)\\*
\nn&=f_{\bS\bV^B}(\bs,\bv^B)\prod_{i\in[d]}\Big(\bbo(w_{0,i}=\rmq_\rms(s_i))\times \prod_{j\in[B]}\bbo(w_{j,i}=\rmq_\rmv^j(v_{j,i}))\Big)\\*
&\qquad\times\Big(\prod_{j\in[B]}\prod_{t\in[n_{j-1}+1,n_j]}\bbo(x_t=\bbo(\rml(\bs,\bv^j,t)\in\calA_t))P_{Y|X}^{\calA_t}(y_t|x_t)\Big)\bbo((\hat{\bs},\hat{\bv}^B)=g(y^{n_B}))\label{def:joint}.
\end{align}
Unless stated otherwise, the probabilities of events are calculated according to the distribution in \eqref{def:joint} or its induced marginal and conditional versions.

It follows that
\begin{align}
\nn&\Pr\{\hat{\bW}_{[0:B]}\neq \bW_{[0:B]}\}\\*
\nn&=\Pr\bigg\{\hat{\bW}_{[0:B]}\neq \bW_{[0:B]}\mathrm{~and~}\|\hat{\bS}-\bS\|_{\infty}>\delta\mathrm{~or~}\exists~j\in[B]:~\|\hat{\bV}_j-\bV\|>\frac{2\delta}{N_j}\bigg\}\\*
&\qquad+\Pr\bigg\{\hat{\bW}_{[0:B]}\neq \bW_{[0:B]}\mathrm{~and~}\|\hat{\bS}-\bS\|_{\infty}\leq\delta,~\forall~j\in[B]:~\|\hat{\bV}_j-\bV\|_{\infty}\leq\frac{2\delta}{N_j}\bigg\}\\
\nn&\leq \Pr\Big\{\|\hat{\bS}-\bS\|_{\infty}>\delta\mathrm{~or~}\exists~j\in[B]:~\|\hat{\bV}_j-\bV\|>\frac{2\delta}{N_j}\Big\}\\*
&\qquad+\Pr\bigg\{\hat{\bW}_{[0:B]}\neq \bW_{[0:B]}\mathrm{~and~}\|\hat{\bS}-\bS\|_{\infty}\leq\delta,~\forall~j\in[B]:~\|\hat{\bV}_j-\bV\|_{\infty}\leq\frac{2\delta}{N_j}\bigg\}\\
&\leq \varepsilon+\Pr\bigg\{\hat{\bW}_{[0:B]}\neq \bW_{[0:B]}\mathrm{~and~}\|\hat{\bS}-\bS\|_{\infty}\leq\delta,~\forall~j\in[B]:~\|\hat{\bV}_j-\bV\|_{\infty}\leq\frac{2\delta}{N_j}\bigg\}\label{useerror4converse}\\
&\leq \varepsilon+\Pr\big\{\hat{\bW}_0\neq \bW_0\mathrm{~and~}\|\hat{\bS}-\bS\|_{\infty}\leq\delta\big\}+\sum_{j\in[B]}\Pr\bigg\{\hat{\bW}_j\neq \bW_j\mathrm{~and~}\|\hat{\bV}_j-\bV\|_{\infty}\leq\frac{2\delta}{N_j}\bigg\}\\
\nn&\leq \varepsilon+\Pr\big\{\exists~i\in[d]:~\hatW_{0,i}\neq W_{0,i}\mathrm{~and~}|\hatS_i-S_i|\leq\delta\big\}\\*
&\qquad+\Pr\big\{\exists~(i,j)\in[d]\times[B]:~\hatW_{j,i}\neq W_{j,i}\mathrm{~and~}N_j|(\hatV_{j,i}-V_{j,i})|\leq 2\delta\big\}\\
&\leq \varepsilon+2d\delta\barM+\Pr\big\{\exists~(i,j)\in[d]\times[B]:~\hatW_{j,i}\neq W_{j,i}\mathrm{~and~}N_j|(\hatV_{j,i}-V_{j,i})|\leq 2\delta\big\}\label{boundaryerror}\\
&\leq \varepsilon+2d\delta\barM+\sum_{j\in[B]}\Pr\big\{\exists~i\in[d]:~\hatW_{j,i}\neq W_{j,i}\mathrm{~and~}N_j|(\hatV_{j,i}-V_{j,i})|\leq 2\delta\big\}\\
&\leq \varepsilon+2d\delta\barM+\sum_{j\in[B]}d\barM_j\frac{4\delta}{N_j}\label{bderror4v}\\
&\leq \varepsilon+2(1+4Bv_+)d\beta\label{use:tilM},
\end{align}
where \eqref{useerror4converse} follows from \eqref{justerror}, \eqref{boundaryerror} follows from the union bound and similar arguments as in \cite{kaspi2018searching} which state that for each $i\in[d]$, the event $\hatW_{0,i}\neq W_{0,i}\mathrm{~and~}|\hatS_i-S_i|\leq\delta$ occurs only when $S_i$ lies near the boundary of a quantized region (see \cite[Appendix B ]{zhou2019twentyq} for a detailed explanation), \eqref{bderror4v} follows similarly to \eqref{boundaryerror}, and \eqref{use:tilM} follows from the definition of $\barM$ and $\barM_j$.

For subsequent analysis, let $\Gamma:[\barM]^d\times\prod_{j\in[B]}[\barM_j]^d\to\calM_{B,d}:=[\barM^d\prod_{j\in[B]}\barM_j^d]$ be an arbitrary one-to-one mapping from $B+1$ $d$-dimensional vectors to an integer. To connect the current problem to data transmission, let $W:=\Gamma(\bW_0,\bW_1,\ldots,\bW_B)$ be a random variable. Similarly, we can define $\hatW=\Gamma(\hat{\bW}_0,\hat{\bW}_1,\ldots,\hat{\bW}_B)$. It follows from \eqref{use:tilM} that
\begin{align}
\Pr\{\hatW\neq W\}
&=\Pr\{\hat{\bW}_{[0:B]}\neq \bW_{[0:B]}\}\\
&\leq \varepsilon+2(1+4B)d\beta
\end{align}
Since we consider uniformly distributed initial location $\bS$ and velocities $\bV^B$, the random variable $W$ is uniformly distributed over $[\calM_{B,d}]$. Given any queries $\calA^{n_B}$, the probability $\Pr\{\hatW\neq W\}$ is the average error probability for a channel coding problem with states at both the encoder and decoder, where at each time $t\in[n_B]$, the channel is given by the query-dependent channel $P_{Y|X}^{f(|\calA_t|)}$. 

\subsubsection{Final steps}
To further lower bound the excess-resolution probability $\varepsilon$, we can use results from finite blocklength information theory, e.g., the meta-converse bound~\cite{polyanskiy2010finite} or the information spectrum method~\cite{TanBook}. Using the non-asymptotic converse bound for channel coding in \cite[Proposition 4.4]{TanBook}, for any $\kappa\in(0,1-\varepsilon-2(1+B)d\beta)$, we have
\begin{align}
\log(|\calM_{B,d}|)
&\leq\sup\bigg\{r\in\bbR_+\Big|\Pr\bigg\{\sum_{t\in[n_B]}\log\frac{P_{Y|X}^{\calA_t}(Y_t|X_t)}{P_Y^{|\calA_t|,f(|\calA_t|)}(Y_t)}\leq r\bigg\}\leq \varepsilon+2(1+4Bv_+)d\beta+\kappa\bigg\}-\log\kappa\label{fbl:converse:final}.
\end{align} 
Note that \eqref{fbl:converse:final} is different from \cite[Proposition 4.4]{TanBook}. In fact, we follow the proof of \cite[Proposition 4.4]{TanBook} with $M$ replaced by $|\calM_{B,d}|$ and $\varepsilon$ replaced by $\varepsilon+2(1+4B)d\beta$ until \cite[Eq. (4.18)]{TanBook}. Then, we use the definition of the so called $\varepsilon$-hypothesis testing divergence~\cite[Eq. (2.9)]{TanBook}. Furthermore, we choose the $Q$ distribution in \cite[Proposition 4.4]{TanBook} as the induced product marginal distribution $\prod_{t\in[n_B]}P_Y^{|\calA_t|,f(|\calA_t|)}$.

It follows from the definitions of $\barM$ and $\barM_j$ that
\begin{align}
\log|\calM_{B,d}|
&=d(\log \barM+\sum_{j\in[B]}\log \barM_j)\\
&=d(\log\barM+\sum_{j\in[B]}\log (2N_jv_+\barM))\\
&=(B+1)d\log\beta-(B+1)d\log\delta+\sum_{j\in[B]}\log (2N_jv_+)\label{messagesize}.
\end{align}
Combining \eqref{fbl:converse:final} and \eqref{messagesize} leads to
\begin{align}
-(B+1)d\log\delta
\nn&\leq\sup_{\calA^{n_B}}\sup\bigg\{r\Big|\Pr\Big\{\sum_{t\in[n_B]}\imath_{\calA_t,f}(X_t;Y_t)\leq r\Big\}\leq \varepsilon+2(1+4Bv_+)d\beta+\kappa\bigg\}\\
&\qquad-(B+1)d\log\beta-\sum_{j\in[B]}\log (2N_jv_+)-\log\kappa.
\end{align}
The proof of Theorem \ref{fbl:converse} is completed.

\subsection{Proof of Theorem \ref{mainresult}}
\label{proof:mainresult}
\subsubsection{Achievability for discrete noise}
\label{sec:dmc}
We first present the achievability proof for discrete noise such that the output alphabet $\calY$ is finite. Consider any $p\in\calP_{\rm{ca}}$. Using Theorem \ref{ach:fbl}, we have that for any $M\in\bbN$ and any ending time points $\bn=(n_1,\ldots.n_B)$ for different slots, the non-adaptive query procedure in Algorithm \ref{procedure:nonadapt} achieves the resolution $\delta=\frac{B+1}{M}$ with excess-resolution probability $\rmP_\rme(\bn,d,\delta)$ satisfying
\begin{align}
\nn&\rmP_\rme\bigg(\bn,d,\frac{B+1}{M}\bigg)
\leq\min\Big\{1,\exp(\zeta(n_1,4,p,v_+,\eta))M^{2d}\mathbb{E}_{P_{XY}^{n_1}}\big[\exp(-\imath_{p,f(p)}(X^{n_1};Y^{n_1}))\big]\Big\}+4n_1\exp(-2(n_1M)^d\eta^2)\\*
&\qquad+ \sum_{j\in[2:B]}\bigg(\min\Big\{1,\exp(\zeta(N_j,3,p,v_+,\eta))M^d\mathbb{E}_{P_{XY}^{N_j}}\big[\exp(-\imath_{p,f(p)}(X^{N_j};Y^{N_j}))\big]\Big\}+4N_j\exp(-2(N_jM)^d\eta^2)\bigg)\label{main:step1},
\end{align}
where $N_j=n_j-n_{j-1}$ for each $j\in[2:B]$. We next analyze each probability term in \eqref{main:step1}. For the first term, we have 
\begin{align}
\nn&\min\Big\{1,\exp(\zeta(n_1,4,p,v_+,\eta))M^{2d}\mathbb{E}_{P_{XY}^{n_1}}[\exp(-\imath_{p,f(p)}(X^{n_1};Y^{n_1}))]\Big\}\\*
&\leq \Pr\bigg\{\exp(\zeta(n_1,4,p,v_+,\eta))M^{2d}\exp(-\imath_{p,f(p)}(X^{n_1};Y^{n_1}))>\frac{1}{n_1}\bigg\}+\frac{1}{n_1}\label{useineq}\\
&=\Pr\big\{\imath_{p,f(p)}(X^{n_1};Y^{n_1})<2d\log M+\zeta(n_1,4,p,v_+,\eta)+\log n_1\big\}+\frac{1}{n_1}\\
&=\Pr\bigg\{\sum_{t\in[n_1]}\imath_{p,f(p)}(X_t;Y_t)<2d\log M+\zeta(n_1,4,p,v_+,\eta)+\log n_1\bigg\}+\frac{1}{n_1}\label{slot1upp},
\end{align}
where \eqref{useineq} follows from the inequality $\min\{1,\mathbb{E}[X]\}\leq \Pr\{X\geq t\}+t$ for any random variable $X$ and positive real number $t\in(0,1)$. Similarly, for each $j\in[2:B]$, we have
\begin{align}
\nn&\min\Big\{1,\exp(\zeta(N_j,3,p,v_+,\eta))M^d\mathbb{E}_{P_{XY}^{N_j}}\big[\exp(-\imath_{p,f(p)}(X^{N_j};Y^{N_j}))\big]\Big\}\\*
&\leq \Pr\bigg\{\sum_{t\in[N_j]}\imath_{p,f(p)}(X_t;Y_t)<d\log M+\zeta(N_j,3,p,v_+,\eta)+\log N_j\bigg\}+\frac{1}{N_j}\label{slotjupp}.
\end{align}

Recall the definitions of $\rmV_p$ in \eqref{def:vp} and $\rmT_p$ in \eqref{def:tp}.
Let $\eta\in\bbR_+$ be chosen as
\begin{align}
\eta=\frac{\log(n_B)}{n_B}.
\end{align}
Let $(\varepsilon_1',\ldots,\varepsilon_B')\in[0,1]^B$ and $\varepsilon'\in[0,1]$ be arbitrary such that $\sum_{j\in[B]}\varepsilon_j'=\varepsilon'$. For each $j\in[B]$, define 
\begin{align}
\varepsilon_j
&:=\varepsilon_j'+4N_j\exp(-2N_j^dM^d\eta^2)+\frac{6T_p}{\sqrt{N_j\rmV_p^3}}\label{def:varepsilonj},
\end{align}
and let $\varepsilon=\sum_{j\in[B]}\varepsilon_j$. Choose $M\in\bbN$ such that
\begin{align}
d\log M
\nn&=\min\bigg\{\frac{n_1C+\sqrt{n_1\rmV_{\varepsilon_1}}\Phi^{-1}(\varepsilon_1')-\zeta(n_1,4,p,v_+,\eta)-\log n_1}{2},\\*
&\qquad\qquad\qquad\min_{j\in[2:B]}\Big(N_jC+\sqrt{N_j\rmV_{\varepsilon_j}}\Phi^{-1}(\varepsilon_j')-\zeta(N_j,3,p,v_+,\eta)-\log N_j\Big)\bigg\}
\label{chooseM:ach}.
\end{align}
Using the Berry-Esseen theorem~\cite{berry1941accuracy,esseen1942liapounoff}, we have that the first term in \eqref{slot1upp} is upper bounded by $\varepsilon_1'+\frac{6T_p}{\sqrt{n_1\rmV_{\varepsilon_1}^3}}$ and for each $j\in[2:B]$, the first term in \eqref{slotjupp} is upper bounded by $\varepsilon_j'+\frac{6T_p}{\sqrt{N_j\rmV_{\varepsilon_j}^3}}$. Thus, the excess-resolution probability satisfies
\begin{align}
\rmP_\rme\bigg(\bn,d,\frac{B+1}{M}\bigg)\leq \varepsilon\label{upperdesired}.
\end{align}

Therefore,
\begin{align}
\nn&-d\log\delta^*(\bn,d,\varepsilon)\\*
&\geq d\log M-d\log(B+1)\\
\nn&=\max_{\substack{(\varepsilon_1,\ldots,\varepsilon_B):\\\sum_{j\in[B]}\varepsilon_j\leq \varepsilon}}
\min\bigg\{\frac{n_1C+\sqrt{n_1\rmV_{\varepsilon_1}}\Phi^{-1}(\varepsilon_1)-\zeta(n_1,4,p,v_+,\eta)-\log n_1+O(1)}{2},\\*
&\qquad\qquad\qquad\qquad\qquad\min_{j\in[2:B]}\Big(N_jC+\sqrt{N_j\rmV_{\varepsilon_j}}\Phi^{-1}(\varepsilon_j)-\zeta(N_j,3,p,v_+,\eta)-\log N_j+O(1)\Big)\bigg\}-d\log(B+1)\label{usetaylor2},
\end{align}
where \eqref{usetaylor2} follows from Taylor expansions of $\Phi^{-1}(\varepsilon_j')$ around $\varepsilon_j'=\varepsilon_j$ and the fact that $\varepsilon_j=\varepsilon_j'+O(\frac{1}{\sqrt{n}})$ that is implied by the following inequalities:
\begin{align}
4n_1\exp(-2n_1^dM^d\eta^2)
&=\exp\bigg(-2n_1^dn_B^{-2}M^d(\log n_B)^2+\log n_1+\log 4\bigg)\\
&=O\left(\exp\left(-\exp\left(\frac{n_1C}{2}+O(\log n_B)\right)\right)+\log n_1\right)\label{main:step3}.
\end{align}
and similarly for each $j\in[2:B]$,
\begin{align}
4N_j\exp(-2N_j^dM^d\eta^2)
&=\exp\bigg(-2N_j^dn_B^{-2}M^d(\log n_B)^2+\log N_j+\log 4\bigg)\\
&=O\left(\exp\left(-\exp\left(N_jC+O(\log n_B)\right)\right)+\log N_j\right).
\end{align}

\subsubsection{Converse Proof}
The converse proof is similar to that of \cite[Theorem 3]{zhou2019twentyq}. Given any sequence of queries $\calA^{n_B}=(\calA_1,\ldots,\calA_{n_B})\in([0,1]^d)^{n_B}$, let
\begin{align}
C_{\calA^{n_B}}&:=\frac{1}{n_B}\sum_{t\in[n_B]}\mathbb{E}[\imath_{\calA_t,f}(X_t;Y_t)],\\
\rmV_{\calA^{n_B}}&:=\frac{1}{n_B}\sum_{t\in[n_B]}\mathrm{Var}[\imath_{\calA_t,f}(X_t;Y_t)],\\
\rmT_{\calA^{n_B}}&:=\frac{1}{n_B}\sum_{t\in[n_B]}\mathbb{E}\Big[\big|\imath_{\calA_t,f}(X_t;Y_t)-\mathbb{E}[\imath_{\calA_t,f}(X_t;Y_t)]\big|^3\Big].
\end{align}

Let $\rmV_-\in\bbR_+$ be chosen such that $\rmV_-\leq \rmV_{\calA^{n_B}}$. For any $r\in\bbR_+$,
 using the Berry-Esseen theorem~\cite{berry1941accuracy,esseen1942liapounoff}, we have that
\begin{align}
\Pr\bigg\{\sum_{t\in[n_B]}\imath_{\calA_t,f}(X_t;Y_t)\leq r\bigg\}
&\geq \Phi\left(\frac{r-n_BC_{\calA^{n_B}}}{\sqrt{n\rmV_{\calA^{n_B}}}}\right)-\frac{6\rmT_{\calA^{n_B}}}{\sqrt{n_B}(\sqrt{\rmV_{\calA^{n_B}}})^3}\\
&\geq  \Phi\left(\frac{r-n_BC_{\calA^{n_B}}}{\sqrt{n\rmV_{\calA^{n_B}}}}\right)-\frac{6\rmT_{\calA^{n_B}}}{\sqrt{n_B}(\sqrt{\rmV_-})^3}.
\end{align}
Thus, for any $\varepsilon\in(0,1)$,
\begin{align}
\nn&\sup\bigg\{r\in\bbR_+:~\Pr\Big\{\sum_{t\in[n_B]}\imath_{\calA_t,f}(X_t;Y_t)\leq r\Big\}\leq \varepsilon\bigg\}\\
&\leq\sup\bigg\{r\in\bbR_+:~\Phi\left(\frac{r-n_BC_{\calA^{n_1}}}{\sqrt{n_B\rmV_{\calA^{n_B}}}}\right)-\frac{6\rmT_{\calA^{n_B}}}{\sqrt{n_B}(\sqrt{\rmV_-})^3}\leq \varepsilon\bigg\}\\
&\leq n_BC_{\calA^{n_B}}+\sqrt{n_B\rmV_{\calA^{n_B}}}\Phi^{-1}\left(\varepsilon+\frac{6\rmT_{\calA^{n_B}}}{\sqrt{n_B}(\sqrt{\rmV_-})^3}\right).
\end{align}
Therefore, choosing $\beta=\frac{1}{\sqrt{n_B}}$ and $\kappa=\frac{1}{n_B}$, we have
\begin{align}
\nn&\sup_{\calA^{n_B}}\sup\bigg\{t\in\bbR_+:~\Pr\Big\{\sum_{t\in[n_B]}\imath_{\calA_t,f}(X_t;Y_t)\leq t\Big\}\leq \varepsilon+2(1+4Bv_+)d\beta+\kappa\bigg\}\\
&\leq \sup_{\calA^{n_B}}\bigg\{n_BC_{\calA^{n_B}}+\sqrt{n_B\rmV_{\calA^{n_B}}}\Phi^{-1}\bigg(\varepsilon+2(1+4Bv_+)d\beta+\kappa+\frac{6\rmT_{\calA^{n_B}}}{\sqrt{n_B}(\sqrt{\rmV_-})^3}\bigg)\bigg\}\\
&=\sup_{\calA^{n_B}:|\calA_t|\in\calP_{\rm{ca}},~t\in[n_B]}\bigg\{n_BC_{\calA^{n_B}}+\sqrt{n_B\rmV_{\calA^{n_B}}}\Phi^{-1}\bigg(\varepsilon+2(1+4Bv_+)d\beta+\kappa+\frac{6\rmT_{\calA^{n_B}}}{\sqrt{n_B}(\sqrt{\rmV_-})^3}\bigg)\bigg\}+O(1)\label{nlarge}\\
&=\sup_{\calA^{n_B}:|\calA_t|\in\calP_{\rm{ca}},~t\in[n_B]}\Bigg(n_BC_{\calA^{n_B}}+\sqrt{n_B\rmV_{\calA^{n_B}}}\Phi^{-1}\left(\varepsilon\right)+O(1)\Bigg)+O(1)\label{taylorphi}\\
&=n_BC+\sqrt{n_B\rmV_{\varepsilon}}\Phi^{-1}(\varepsilon)+O(1),\label{usedefs}
\end{align}
where \eqref{nlarge} follows from \cite[Lemma 49]{polyanskiy2010thesis}, \eqref{taylorphi} follows from the Taylor expansion for $\Phi^{-1}(\cdot)$ at around $\varepsilon$,  and \eqref{usedefs} follows from the definitions of $C$ in \eqref{def:c} and $\rmV_\varepsilon$ in \eqref{def:v}.

Therefore, using Theorem \ref{fbl:converse}, we have
\begin{align}
-(B+1)d\log\delta^*(\bn,d,\varepsilon)
\leq n_BC+\sqrt{n_B\rmV_{\varepsilon}}\Phi^{-1}(\varepsilon)+O(\log n_B).
\end{align}

\section{Conclusion}
\label{sec:conclusion}

We addressed optimal search for a moving target with unknown initial location and piecewise constant velocity model over the unit cube of a finite dimension and derived bounds on the performance of optimal non-adaptive query procedures. Our bounds are tight first-order asymptotically when the number of queries in each time slot satisfies mild conditions and imply that cold restart search is strictly suboptimal. When the target moves with a constant velocity, our bounds are tight second-order asymptotically and provide approximation to the finite query performance of optimal non-adaptive querying. In this case, our results imply an interesting phase transition phenomenon for the excess-resolution probability as a function of the resolution decay rate.

Future directions include the following. Firstly, it would be worthwhile to relax the ``torus'' constraint in Eq. \eqref{location} on the moving target and study the fundamental limit of optimal search strategies under more practical boundary assumptions. Secondly, one can generalize our piecewise velocity model to account for more practical settings by assuming that the velocity of the target is a smooth function of time with bounded derivatives or tolerating random noise and perturbation in the trajectory from piecewise constant velocity models. The main challenge will be to be able to analyze the set of all possible trajectories for such a setting in the achievability proof of Theorem 4. In the converse part, novel ideas will be required in order to relate the search problem to channel coding. 
Thirdly, in this paper and previous studies on 20 questions estimation, e.g.,~\cite{kaspi2018searching,chung2018unequal,zhou2019twentyq}, the query set is allowed to be arbitrary. However, in practical search problems such as beam alignment~\cite{chiu2019beam}, the query set is usually connected. Nonetheless, it would be interesting to generalize our results in this paper and companion papers~\cite{zhou2019twentyq,zhou2021resolution} to the case of hierarchical and dyadic query sets as in~\cite{chiu2021search}. Fourthly, in Algorithm \ref{procedure:nonadapt}, we use random sampling, which enables us to derive theoretical benchmarks. However, such an algorithm suffers from high computational complexity due to the need for exhaustive decoding over all possible trajectories, which renders it impractical. It would be fruitful to investigate the existence of a low complexity deterministic algorithm that achieves or approximates our theoretical benchmarks. Finally, in this paper, we focused on non-adaptive query procedures due to its time-efficiency. However, adaptive query procedures usually yield superior performance~\cite{zhou2019twentyq,lalitha2018improved}. Thus, one can study adaptive query procedures and explore the potential benefit of adaptivity. In this direction, one might use the concepts in~\cite{Kartik2022active,chiu2016sequential} to design and analyze adaptive query procedures.

\appendix 
\subsection{Equivalence between Maximal Mutual Information Density Decoding and Nearest Neighbor Decoding}
\label{just:nnd}

It suffices to justify for the decoding for the first time slot since the same decoding method is used in subsequent time slots. Fix $n_1\in\bbN$. Consider a query-dependent BSC with parameter $\zeta\in(0,1)$~(cf. Definition \ref{def:mdBSC}). For any $p\in(0,1)$, let
\begin{align}
\beta(p)=p(1-f(p))+(1-p)f(p).
\end{align}
For any $(x^{n_1},y^{n_1})\in\{0,1\}^{2n}$, the mutual information density is given by
\begin{align}
\imath_{p,f(p)}(x^{n_1};y^{n_1})
\nn&=\sum_{t\in[n_1]}\Big(\bbo(x_t\neq y_t)\log(\zeta f(p))+\bbo(x_t=y_t)\log(1-\zeta f(p))\\*
&\qquad\qquad-\bbo(y_t=1)\log(\beta(p))-\bbo(y_t=0)\log(1-\beta(p))\Big).
\end{align}
It follows that
\begin{align}
\nn&\argmax_{\bar{\bw}\in\calB_{n_1,M}}\imath_{p,f(p)}\left(x^{n_1}(\bar{\bw}^{n_1});y^{n_1}\right)\\
&=\argmax_{\bar{\bw}\in\calB_{n_1,M}}\bigg(\sum_{t\in[n_1]}\Big(\bbo\Big(x_t(\bar{\bw}_t)\neq y_t\Big)\log(\zeta f(p))+\bbo\Big(x_t(\bar{\bw}_t)=y_t\Big)\log(1-\zeta f(p))\bigg)\\
&=\argmax_{\bar{\bw}\in\calB_{n_1,M}}\bigg(\left\|y^{n_1}-x^{n_1}(\bar{\bw}^{n_1})\right\|^2\log(\zeta f(p))+\left(n-\left\|y^{n_1}-x^{n_1}(\bar{\bw}^{n_1})\right\|^2\right)\log(1-\zeta f(p))\bigg)\\
&=\argmin_{\bar{\bw}\in\calB_{n_1,M}}\left\|y^{n_1}-x^{n_1}(\bar{\bw}^{n_1})\right\|^2\label{nnd:bsc},
\end{align}
where \eqref{nnd:bsc} follows since $\zeta f(p)\leq \frac{1}{2}$ from our assumption below Definition \ref{def:mdBSC}. Thus, we have justified that maximal mutual information density decoding is equivalent to nearest neighbor decoding for a query-dependent BSC. 

Now consider a query-dependent AWGN channel. From Definition \ref{def:mdAWGN}, for any $p\in(0,1)$ and $(x^{n_1},y^{n_1})\in\{0,1\}^{n_1}\times\bbR^{n_1}$, the mutual information density is
\begin{align}
\imath_{p,f(p)}(x^{n_1};y^{n_1})
&=\sum_{t\in[n_1]}\Big(-\frac{(y_t-x_t)^2}{2(f(p)\sigma)^2}-\log P_Y^{p,(f(p))}(y_t)\Big).
\end{align}
Thus, given any $y^{n_1}$,
\begin{align}
\nn
&\argmax_{\bar{\bw}\in\calB_{n_1,M}}\imath_{p,f(p)}\left(x^{n_1}(\bar{\bw}^{n_1});y^{n_1}\right)\\
\nn&=\argmax_{\bar{\bw}\in\calB_{n_1,M}}\sum_{t\in[n_1]}\bigg(-\frac{(y_t-x_t(\bar{\bw}_t))^2}{2(f(p)\sigma)^2}-\log P_Y^{p,(f(p))}(y_t)\bigg)\\
&=\argmin_{\bar{\bw}\in\calB_{n_1,M}}\left\|y^{n_1}-x^{n_1}(\bar{\bw}^{n_1})\right\|^2.
\end{align}
Thus, we obtain equivalence between maximal mutual information density decoding and nearest neighbor decoding for the query-dependent AWGN channel.

\subsection{Justification of \eqref{prob_atyp_noise}}
\label{justify:prob_noise}
Consider any Gaussian random variable $Z$ with mean $0$ and variance $\sigma^2$. For any $\theta$ such that $2\theta\sigma^2<1$, we have
\begin{align}
\mathbb{E}[\exp(\theta Z^2)]
&=\int_{-\infty}^{\infty}\frac{1}{\sqrt{2\sigma^2}}\exp\left(-\frac{z^2}{2\sigma^2}\right)\exp(\theta z^2) \rmd z\\
&=\int_{-\infty}^{\infty}\frac{1}{\sqrt{2\sigma^2}}\exp\left(-\frac{z^2}{2\sigma^2}+\theta z^2\right)\rmd z\\
&=\int_{-\infty}^{\infty}\frac{1}{\sqrt{2\sigma^2}}\exp\left(-\frac{(1-2\sigma^2\theta)z^2}{2\sigma^2}\right)\rmd z\\
&=\frac{1}{\sqrt{1-2\theta\sigma^2}}\int_{-\infty}^{\infty}\sqrt{\frac{1-2\theta\sigma^2}{2\sigma^2}}\exp\left(-\frac{(1-2\theta\sigma^2)z^2}{2\sigma^2}\right)\rmd z\\
&=\frac{1}{\sqrt{1-2\theta\sigma^2}},
\end{align}
and for $\theta\geq \frac{1}{2\sigma^2}$, we have
\begin{align}
\mathbb{E}[\exp(\theta Z^2)]=\infty.
\end{align}

For any $(\theta,\sigma)\in\bbR_+^3$, let
\begin{align}
I(\theta,\sigma^2)
&:=2\theta\sigma^2-\log\big(\mathbb{E}[\exp(\theta Z_1^2)])\big)=\left\{
\begin{array}{cc}
2\theta\sigma^2+\frac{1}{2}\log(1-2\theta\sigma^2)&\mathrm{if~}\theta<\frac{1}{2\sigma^2}\\
-\infty&\mathrm{otherwise}
\end{array}
\right.
\end{align}
Then, for any $\theta\in\bbR_+$,
\begin{align}
\Pr\bigg\{\frac{1}{n_1}\sum_{t\in[n_1]}Z_t^2>2\sigma^2\bigg\}
&=\Pr\bigg\{\sum_{t\in[n_1]}\theta Z_t^2>2 n_1\theta\sigma^2\bigg\}\\
&=\Pr\bigg\{\exp\Big(\sum_{t\in[n_1]}\theta Z_t^2\Big)>\exp(2 n_1\theta\sigma^2)\bigg\}\\
&\leq \frac{\mathbb{E}\Big[\exp\big(\sum_{t\in[n_1]}\theta Z_t^2\big)\Big]}{\exp(2 n_1\theta\sigma^2)}\\
&=\exp\Big(-2n_1\theta \sigma^2+n_1\log\big(\mathbb{E}[\exp(\theta Z_1^2)]\big)\Big)\\
&=\exp(-n_1I(\theta,\sigma^2))\label{tocombine1}.
\end{align}

For any $\theta\in(0,\frac{1}{2\sigma^2})$, the first derivative of $I(\theta,\sigma^2)$ with respect to $\theta$ is
\begin{align}
\frac{\partial I(\theta,\sigma^2)}{\partial \theta}
&=2\sigma^2-\frac{\sigma^2}{1-2\theta\sigma^2}
\end{align}
Thus, 
\begin{align}
\max_{\theta\in\bbR_+}I(\theta,\sigma^2)
&=\max_{\theta\in\bbR_+:\theta<\frac{1}{2\sigma^2}}I(\theta,\sigma^2)=I\left(\frac{1}{4\sigma^2},\sigma^2\right)=\frac{1-\log 2}{2}>0\label{tocombine2}.
\end{align}
The justification of \eqref{prob_atyp_noise} is completed by combining \eqref{tocombine1} and \eqref{tocombine2}.

\section*{Acknowledgments}
The authors acknowledge two anonymous reviewers for helpful comments and suggestions, which help imoprimprove the quality of the current manuscript.

\bibliographystyle{IEEEtran}
\bibliography{IEEEfull_lin}

\begin{thebibliography}{10}
\providecommand{\url}[1]{#1}
\csname url@samestyle\endcsname
\providecommand{\newblock}{\relax}
\providecommand{\bibinfo}[2]{#2}
\providecommand{\BIBentrySTDinterwordspacing}{\spaceskip=0pt\relax}
\providecommand{\BIBentryALTinterwordstretchfactor}{4}
\providecommand{\BIBentryALTinterwordspacing}{\spaceskip=\fontdimen2\font plus
\BIBentryALTinterwordstretchfactor\fontdimen3\font minus
  \fontdimen4\font\relax}
\providecommand{\BIBforeignlanguage}[2]{{%
\expandafter\ifx\csname l@#1\endcsname\relax
\typeout{** WARNING: IEEEtran.bst: No hyphenation pattern has been}%
\typeout{** loaded for the language `#1'. Using the pattern for}%
\typeout{** the default language instead.}%
\else
\language=\csname l@#1\endcsname
\fi
#2}}
\providecommand{\BIBdecl}{\relax}
\BIBdecl

\bibitem{zhouicassp2020}
L.~{Zhou} and A.~O. {Hero}, ``Resolution limits of 20 questions search
  strategies for moving targets,'' in \emph{IEEE ICASSP}, 2021.

\bibitem{polyanskiy2010finite}
Y.~Polyanskiy, H.~V. Poor, and S.~Verd\'u, ``Channel coding rate in the finite
  blocklength regime,'' \emph{IEEE Trans. Inf. Theory}, vol.~56, no.~5, pp.
  2307--2359, 2010.

\bibitem{TanBook}
V.~Y.~F. Tan, ``Asymptotic estimates in information theory with non-vanishing
  error probabilities,'' \emph{{Foundations and Trends$\,$\textregistered $ $
  in Communications and Information Theory}}, vol.~11, no. 1--2, pp. 1--184,
  2014.

\bibitem{ZhouBook}
L.~Zhou and M.~Motani, ``Finite blocklength lossy source coding for discrete
  memoryless sources,'' \emph{Foundations and Trends$\,$\textregistered $ $ in
  Communications and Information Theory}, vol.~20, no.~3, pp. 157--389, 2023.

\bibitem{wiki20question}
\BIBentryALTinterwordspacing
Wikipedia. (2020, 05) Twenty questions. [Online]. Available:
  \url{https://en.wikipedia.org/wiki/Twenty_Questions}
\BIBentrySTDinterwordspacing

\bibitem{renyi1961problem}
A.~R{\'e}nyi, ``On a problem of information theory,'' \emph{MTA Mat. Kut. Int.
  Kozl. B}, vol.~6, pp. 505--516, 1961.

\bibitem{ulam1991adventures}
S.~M. Ulam, \emph{Adventures of a Mathematician}.\hskip 1em plus 0.5em minus
  0.4em\relax Univ. of California Press, 1991.

\bibitem{zhou2019twentyq}
L.~Zhou and A.~O. Hero, ``Resolution limits for the noisy non-adaptive 20
  questions problem,'' \emph{IEEE Trans. Inf. Theory}, vol.~67, no.~4, pp.
  2055--2073, 2021.

\bibitem{tsiligkaridis2014collaborative}
T.~Tsiligkaridis, B.~M. Sadler, and A.~O. Hero, ``Collaborative 20 questions
  for target localization,'' \emph{IEEE Trans. Inf. Theory}, vol.~60, no.~4,
  pp. 2233--2252, 2014.

\bibitem{jedynak2012twenty}
B.~Jedynak, P.~I. Frazier, and R.~Sznitman, ``Twenty questions with noise:
  Bayes optimal policies for entropy loss,'' \emph{J. Appl. Probab.}, vol.~49,
  no.~1, pp. 114--136, 2012.

\bibitem{rajan2015bayesian}
P.~Rajan, W.~Han, R.~Sznitman, P.~Frazier, and B.~Jedynak, ``Bayesian multiple
  target localization,'' \emph{J. Mach. Learn. Res.}, vol.~37, pp. 1945--1953,
  2015.

\bibitem{pelc2002searching}
A.~Pelc, ``Searching games with errors---fifty years of coping with liars,''
  \emph{Theor. Comput. Sci.}, vol. 270, no. 1-2, pp. 71--109, 2002.

\bibitem{chiu2016sequential}
S.-E. Chiu and T.~Javidi, ``Sequential measurement-dependent noisy search,'' in
  \emph{IEEE ITW}, 2016, pp. 221--225.

\bibitem{tsiligkaridis2015decentralized}
T.~Tsiligkaridis, B.~M. Sadler, and A.~O. Hero, ``On decentralized estimation
  with active queries,'' \emph{IEEE Trans. Signal Process.}, vol.~63, no.~10,
  pp. 2610--2622, 2015.

\bibitem{kaspi2018searching}
Y.~Kaspi, O.~Shayevitz, and T.~Javidi, ``Searching with measurement dependent
  noise,'' \emph{IEEE Trans. Inf. Theory}, vol.~64, no.~4, pp. 2690--2705,
  2018.

\bibitem{chung2018unequal}
H.~W. Chung, B.~M. Sadler, L.~Zheng, and A.~O. Hero, ``Unequal error protection
  querying policies for the noisy 20 questions problem,'' \emph{IEEE Trans.
  Inf. Theory}, vol.~64, no.~2, pp. 1105--1131, 2018.

\bibitem{variani2015non}
E.~Variani, K.~Lahouel, A.~Bar-Hen, and B.~Jedynak, ``Non-adaptive policies for
  20 questions target localization,'' in \emph{IEEE ISIT}, 2015, pp. 775--778.

\bibitem{csiszar2011information}
I.~Csisz\'ar and J.~K{\"o}rner, \emph{Information Theory: Coding Theorems for
  Discrete Memoryless Systems}.\hskip 1em plus 0.5em minus 0.4em\relax
  Cambridge University Press, 2011.

\bibitem{cover2012elements}
T.~M. Cover and J.~A. Thomas, \emph{Elements of information theory}.\hskip 1em
  plus 0.5em minus 0.4em\relax John Wiley \& Sons, 2012.

\bibitem{wolfobook}
J.~Wolfowitz, ``Coding theorems of information theory,'' \emph{New
  York:Springer-Verlag}, 2016.

\bibitem{lalitha2018improved}
A.~Lalitha, N.~Ronquillo, and T.~Javidi, ``Improved target acquisition rates
  with feedback codes,'' \emph{IEEE J. Sel. Topics Signal Process.}, vol.~12,
  no.~5, pp. 871--885, 2018.

\bibitem{zhou2021resolution}
L.~Zhou, L.~Bai, and A.~O. Hero, ``Resolution limits of non-adaptive 20
  questions search for multiple targets,'' \emph{IEEE Trans. Inf. Theory},
  vol.~68, no.~8, pp. 4964--4982, 2022.

\bibitem{el2011network}
A.~El~Gamal and Y.-H. Kim, \emph{Network Information Theory}.\hskip 1em plus
  0.5em minus 0.4em\relax Cambridge University Press, 2011.

\bibitem{tantomamichel2015}
V.~Y.~F. Tan and T.~Tomamichel, ``The third-order term in the normal
  approximation for the {AWGN} channel,'' \emph{IEEE Trans. Inf. Theory},
  vol.~61, no.~5, pp. 2430--2438, 2015.

\bibitem{berry1941accuracy}
A.~C. Berry, ``The accuracy of the {Gaussian} approximation to the sum of
  independent variates,'' \emph{Trans. Am. Math. Soc.}, vol.~49, no.~1, pp.
  122--136, 1941.

\bibitem{esseen1942liapounoff}
C.-G. Esseen, \emph{On the {Liapounoff} limit of error in the theory of
  probability}.\hskip 1em plus 0.5em minus 0.4em\relax Almqvist \& Wiksell,
  1942.

\bibitem{kostinajscc}
V.~Kostina and S.~Verd{\'u}, ``Lossy joint source-channel coding in the finite
  blocklength regime,'' \emph{IEEE Trans. Inf. Theory}, vol.~59, no.~5, pp.
  2545--2575, 2013.

\bibitem{tan2014state}
M.~Tomamichel and V.~Y.~F. Tan, ``Second-order coding rates for channels with
  state,'' \emph{IEEE Trans. Inf. Theory}, vol.~60, no.~8, pp. 4427--4448,
  2014.

\bibitem{bouleau1994numerical}
N.~Bouleau and D.~Lepingle, \emph{Numerical methods for stochastic
  processes}.\hskip 1em plus 0.5em minus 0.4em\relax John Wiley \& Sons, 1994,
  vol. 273.

\bibitem{polyanskiy2010thesis}
Y.~Polyanskiy, ``Channel coding: Non-asymptotic fundamental limits,'' Ph.D.
  dissertation, Department of Electrical Engineering, Princeton University,
  2010.

\bibitem{chiu2019beam}
S.-E. Chiu, N.~Ronquillo, and T.~Javidi, ``Active learning and {CSI}
  acquisition for mmwave initial alignment,'' \emph{IEEE J. Sel. Areas
  Commun.}, vol.~37, no.~11, pp. 2474--2489, 2019.

\bibitem{chiu2021search}
S.-E. Chiu and T.~Javidi, ``Low complexity sequential search with
  size-dependent measurement noise,'' \emph{IEEE Trans. Inf. Theory}, vol.~67,
  no.~9, pp. 5731--5748, 2021.

\bibitem{Kartik2022active}
D.~Kartik, A.~Nayyar, and U.~Mitra, ``Fixed-horizon active hypothesis
  testing,'' \emph{IEEE Trans. Autom. Control}, vol.~67, no.~4, pp. 1882--1897,
  2022.

\end{thebibliography}
\end{document}